\documentstyle[11pt,epsfig]{article}

\newcommand \be{\begin{equation}}
\newcommand \ba{\begin{eqnarray}}
\newcommand \ee{\end{equation}}
\newcommand \ea{\end{eqnarray}}

\begin{document}

\begin{center}
{\LARGE Non-Parametric Analyses of Log-Periodic Precursors to
Financial Crashes}
\end{center}
\bigskip
\begin{center}
{\large Wei-Xing Zhou {\small$^{\mbox{\ref{igpp}}}$} and Didier
Sornette {\small$^{\mbox{\ref{igpp},\ref{ess},\ref{lpec}}}$}}
\end{center}
\bigskip
\begin{enumerate}
\item Institute of Geophysics and Planetary Physics, University of
California, Los Angeles, CA 90095\label{igpp}

\item Department of Earth and Space Sciences, University of
California, Los Angeles, CA 90095\label{ess}

\item Laboratoire de Physique de la Mati\`ere Condens\'ee, CNRS UMR
6622 and Universit\'e de Nice-Sophia Antipolis, 06108 Nice Cedex
2, France\label{lpec}

\end{enumerate}

\begin{abstract}

We apply two non-parametric methods to test further the hypothesis
that log-periodicity characterizes the detrended price trajectory
of large financial indices prior to financial crashes or strong
corrections. The analysis using the so-called $(H,q)$-derivative
is applied to seven time series ending with the October 1987
crash, the October 1997 correction and the April 2000 crash of the
Dow Jones Industrial Average (DJIA), the Standard \& Poor 500 and
Nasdaq indices. The Hilbert transform is applied to two detrended
price time series in terms of the $\ln(t_c-t)$ variable, where
$t_c$ is the time of the crash. Taking all results together, we
find strong evidence for a universal fundamental log-frequency $f
= 1.02 \pm 0.05$ corresponding to the scaling ratio $\lambda =
2.67 \pm 0.12$. These values are in very good agreement with those
obtained in past works with different parametric techniques.

\end{abstract}

\section{Introduction}
\label{sec:intro}

Several recent analyses
\cite{JSoutlier1,JSoutlier2,SorJoh01,SorPNAS,Mansilla,Lamper}
have shown that large financial crashes
are ``outliers'' or ``kings'' \cite{king} with statistically different
properties than the rest of the population of price returns.
A proposed mechanism for understanding this ``king'' effect
involves collective interactions between agents leading
to a cascade of amplifications,
which may result from several origins
\cite{JohSorLed99,CriCrash99,CriCrash00,SorAndersen,bookcrash,Lamper,Idesor}.

Associated with this ``king'' effect, log-periodic patterns
\cite{SorDSI} in financial price time series have been found to be
precursory signatures of large crashes or large drawdowns
\cite{SJB,Feigen1,SJ96,Vande,Van2,Drozdz,Van3,Feigen2,Nasdaq,Quantfinancefeigen,emergent,Bothmer}
(see also \cite{SorJoh01,bookcrash} and references therein). Most
of the analyses have been of a parametric nature, based on
formulas involving combinations of powers laws and of log-periodic
functions of the type $\cos [\omega \ln (t_c-t)]$
\cite{SJ96,model,anti}, where $t_c$ is a ``critical'' time at or
close to the crash.  A first attempt at providing a non-parametric
test of log-periodicity in financial time series was based on a
spectral Lomb analysis \cite{Press} of the oscillatory residuals
to a preliminary power law fit to the time series
\cite{CriCrash99,CriCrash00,SorJoh01}.

Here, we propose to extend the search for non-parametric
signatures of log-periodicity. For this, we introduce two
non-parametric techniques and apply them to seven financial time
series culminating in a crash or a strong correction. The first
non-parametric method is called the generalized $q$-analysis or
$(H,q)$-analysis and was introduced in Ref.~\cite{Zhou02c} in the
analogous context of the detection of log-periodic oscillations in
precursory signals to material failures. The second non-parametric
method uses the Hilbert transform to construct the local
cumulative phase of the oscillatory component in the time series.
By the new lights provided by these analyses that are cast on the
difficult empirical problem of detecting reliable precursory
patterns to financial crashes or to strong corrections, we hope to
provide what we believe are strong additional evidence for the
reality of log-periodicity in price trajectories. Compared with a
similar analysis of material failures using the $(H,q)$-analysis
\cite{Zhou02c}, we find that log-periodicity seems to be a
significantly stronger signal in financial time series than it is
for rupture.

\section{The generalized $(H,q)$-analysis}
\label{s:HqA}

\subsection{Definitions}

The so-called $q$-analysis is
based on the concept of a $q$-derivative, the
inverse of the Jackson $q$-integral \cite{Jacksonqder}, which is
the natural tool \cite{Erzan,ErzEck} for describing  discrete
scale invariance \cite{SorDSI}.
The $q$-derivative of an arbitrary function $f(x)$ for
any $q\in (0,1)\cup (1,\infty)$ is defined as
\begin{equation}
D_q f(x) = \frac{f(x)-f(qx)}{(1-q)x}~, \label{Eq:qD}
\end{equation}
where $x$ is the distance to the critical point. If the signal is
a price $p(t)$ function to time $t$, then $x=t_c-t$, where $t_c$
is the critical time, that is, the time of the culmination of the
bubble, which is in general close to the time of the crash
\cite{CriCrash00,JohSorLed99}.
The case $q \to 1$
recovers the normal definition of derivative. It follows that
\begin{equation}
D_{1/q} f(x) = D_{q} f(x/q). \label{Eq:map}
\end{equation}
It is thus enough to study $D_q f(x)$ for $q \in (0,1)$ to derive
its values for all $q$'s.

The necessary and sufficient condition for a function $f(x)$ to be
homogeneous of order $\psi$ is
\begin{equation}
D_q f(x) = \frac{q^\psi-1}{q-1} \frac{f(x)}{x}~.
\label{Eq:HomoCond}
\end{equation}
This expression suggests the introduction of a generalized
$q$-derivative that we call $(H,q)$-derivative, such that the
dependence in $x$ of $D_q^H f(x)$ disappears for homogeneous
functions, for the choice $H=\psi$. Consider therefore the
following definition
\begin{equation}
D_q^H f(x)
\stackrel{\triangle}{=} \frac {f(x)-f(qx)}{[(1-q)x]^H}~,
\label{Eq:HqD}
\end{equation}
such that $D_q^{H=1} f(x)$ recovers the standard $q$-derivative
$D_q f(x)$. For a power law function $f(x) = B x^m$, $D_q^{H=m} [B
x^m] = B (1-q^m)/(1-q)^m$ is constant. For a statistically
homogeneous function $f(x) \stackrel{d}{=} B x^m$, $D_q^{H=1} f(x)
\stackrel{d}{=}$ constant, where the symbol $\stackrel{d}{=}$
means equivalence in distribution.

The generalized $(H,q)$-derivative has two control parameters: the
discrete scale factor $q$ devised to characterize the log-periodic
structure and the exponent $H$ introduced to account for a
possible power law dependence, i.e., to correct for
the existence of trends in log-log plots.

Let us consider the following
log-periodic function of $x = (t_c-t)/\tau$:
\begin{equation}
y(x) = A - B\tau^\beta x^\beta  + C\tau^\beta x^\beta
\cos\left(\omega\ln{x} \right)~, \label{Eq:Sor}
\end{equation}
where the phase in the cosine $\phi = \omega \ln{\tau}$ has been
scaled away by a change of time scale. In the following, we shall
refer to $\omega$ as the angular log-frequency and shall also use
the log-frequency defined as \be f = {\omega \over 2 \pi}~.
\label{deff} \ee The $(H,q)$-derivative of $y(x)$ is
\begin{equation}
D_q^H y(x) = x^{\beta -H} [B' + C'g(x)]. \label{Eq:DqHlogcos}
\end{equation}
where $B' =-{ B\tau^\beta (1-q^\beta ) \over (1-q)^H}$,
$C'={C\tau^\beta  \over (1-q)^H}$ and
\begin{eqnarray}
g(x) &=& \cos(\omega \ln x)-q^\beta \cos(\omega
\ln{q x})\\
&=& C_1 \cos(\omega \ln{x}) + C_2\sin(\omega\ln{x}),
\label{Eq:bracket}
\end{eqnarray}
where $C_1=[1 - q^\beta \cos(\omega \ln{q})]>0$ and
$C_2=\sin(\omega\ln{q})$. In principle, we could optimize the value
$q$ to adapt it to the angular log-frequency $\omega$ to ensure $C_1=0$,
so that $g(x)$ becomes a pure sinus: a study of the phase of the log-periodic
oscillations as a function of $q$ can in principle get access to $\omega$.
We have not tried this approach as it seems very sensitive to noise.

Varying $\beta$, we find that the analysis in terms of the
$(H,q)$-derivative of $y(x)$ given by (\ref{Eq:Sor}) gives
essentially the same results for the highest peaks $P_N(H,q)$ and
their associated log-frequencies $f(H,q)$ when varying $\beta$ at
fixed $\omega=2\pi f$ but gives very different and volatile
results when varying $\omega$ at fixed $\beta$. This shows the
value of the $(H,q)$-analysis which is a robust detector of the
log-frequency but not of the power law trend.

\subsection{Effect of errors in the critical time $t_c$}
\label{ss:Transform}

An important issue is the determination of the critical time $t_c$
entering in the definition of the scaling parameter $x = t_c-t$,
such that the time series is (discretely) scale invariant with
respect to scale transformation performed on $x$. Indeed,
according to the theory of rational bubbles and crashes
\cite{CriCrash00,JohSorLed99,SorAndersen}, the crash is not a
certain deterministic event but is described by its hazard rate
$h(t)$, which is the probability per unit time that the market
crashes conditioned on the fact that it has not crashed yet. In
this framework, the critical $t_c$ is the end of the bubble and
coincides with the peak of the crash hazard rate. In practice,
this means that the time of the crash, if it occurs, is expected
to be close to and slightly smaller than $t_c$. Thus, knowing the
time of the crash does not provide a perfect measurement of $t_c$.
Therefore, there is a fundamental and unavoidable uncertainty in
the determination of the value of $t_c$ entering the definition of
$x$. This uncertainty is a severe limitation for parametric
methods of detections of log-periodicity. In contrast, our
previous work \cite{Zhou02c} has shown that the $(H,q)$-derivative
is rather insensitive to an error in the estimation of $t_c$ up to
times of the order of the error made on the determination of $t_c$
(see figure 2 of \cite{Zhou02c}). Technically, this can be seen
from the following derivation.

An error in $t_c$ amounts to a spurious translation of the
coordinate system $(x,y)$ away from the origin $(0, 0)$ to a new
origin $(x_0,y_0)$. This defined the following coordinate
transformation $(x,y) \to ({\bar{x}}, {\bar{y}})$:
\begin{equation}
    \left\{ {\begin{array}{*{1}c}
      x \to {\bar{x}} = x-x_0  \\
      y \to {\bar{y}} = y-y_0, \\
    \end{array} } \right.
\end{equation}
where $y_0=y(x_0)$. According to this transform,
${\bar{y}}(\bar{x})=y(x)-y_0 = y(\bar{x}+x_0)-y_0$ and
${\bar{y}}(q\bar{x})=y(q\bar{x}+x_0)-y_0 = y(qx-qx_0+x_0)$. Thus,
the $(H,q)$-derivative is
\begin{eqnarray}
D_q^H \bar{y}(\bar{x}) ={{B\bar{q}^\beta
-B+C\cos(\omega\ln{x})-C\bar{q}^\beta
\cos[\omega\ln(\bar{q}x)]}\over{(1-q)^H(x-x_0)^Hx^{-m}}}
\label{njgnw}
\end{eqnarray}
where $\bar{q}=q+{(1-q)x_0\over{x}}$.

Consider two values $x_{n+1}
>  x_n$ of the control parameter $x$
corresponding to successive local extrema of the log-periodic
function, such that $\cos[\omega\ln(\bar{q}_nx_n)] =
\cos[\omega\ln(\bar{q}_{n+1}x_{n+1})]$, that is, \be
\ln(\bar{q}_nx_n) - \ln(\bar{q}_{n+1}x_{n+1}) = 2k\pi/\omega
\label{Eq:TranXbar}~. \ee Therefore,
\begin{equation}
{x_{n+1}-qx_0+x_0 \over x_{n}-qx_0+x_0} = e^{2k\pi \over \omega}
\end{equation}
For $x_{n+1} > x_n \gg x_0 $, we obtain \be \ln{x_{n+1}} -
\ln{x_n} = 2k\pi/\omega~, \label{Eq:TranX} \ee that is, the
translational error in the $x$ variable has a negligible impact on
the determination of the angular log-frequency $\omega$ defined by
(\ref{Eq:Sor}) and recovered by the measure (\ref{Eq:TranX}), as
long as the analysis is not too close to the critical point. In
contrast, consider for instance $x_1 < x_2$ comparable to $x_0$
and which satisfy Eq.~(\ref{Eq:TranX}) with $k = 1$. In the
translated coordinate system, the phase difference between these
two points of the term $C\bar{q}^\beta \cos[\omega\ln(\bar{q}x)]$
in the $(H,q)$-derivative of $y(x)$ given by (\ref{njgnw}) is
given by
\begin{equation}
\Delta \phi = \omega \ln {x_2-qx_0+x_0 \over x_1-qx_0+x_0}~.
\label{Eq:DeltaPhi}
\end{equation}
Since $0<q<1$, $0 < \Delta \phi < 2\pi$, which implies
that the ``period'' near $x_0$ is stretched by the translation of $x$
to the uncorrect critical value. In this case, the
log-periodic oscillations disappear for $x$ of the order of and
less than $x_0$ (see figure 2 of \cite{Zhou02c}).

Suppose that the log-periodic oscillations for times $t_c - t <
t_c -t_{\max}$ are not well-determined or unobservable. Then, for
a given $q$, the $(H,q)$-derivative is not well-determined for \be
t > t_c - {t_c-t_{\max} \over q}~. \label{Eq:ttt} \ee The smaller
$q$ is, the larger is the time interval in which the
$(H,q)$-derivative is not determined. This explains our results
below that small $q$'s do not give good results and tend to
over-emphasize slow log-periodic oscillations, hence the existence
of the so-called and spurious ${\mathbf{B}}$ platform at very low
log-frequency. The existence of spurious low log-frequencies is a
general phenomenon which may derive from a variety of origins
\cite{fmp}.

In the sequel, we shall not attempt to optimize our analysis of
$t_c$ and shall fix $t_c$ to the time of the crash or strong
correction ending each bubble, keeping in mind that one should
thus expect a deterioration of the searched signal close to the
end of the time series. In practice, we have not found this to be
a limitation for the present purpose of establishing the existence
of the log-periodic patterns. However, the situation is not as
bright when attempting to use this analysis in a prediction mode,
that is, by scanning $t_c$ with an incomplete time series ending
earlier than the crash, as discussed in \cite{Zhou02c}.

\subsection{Procedure and selection criteria}
\label{ss:Criteria}

Our analysis of a given time series consists in the following steps.
\begin{enumerate}
\item Read the critical time $t_c$ as given by the time of the crash or of
the strong correction and
transform the time series $p(t)$ into a function $y(x)=p(t)$
with $x=t_c-t$.

\item Fix $q$ and $H$ to some arbitrary value and construct the
corresponding $(H,q)$-derivative $D_q^H p(\tau)$ of the function
$y(x)$.

\item Perform a spectral Lomb analysis \cite{Horne,Press} of $D_q^H p(\tau)$
in order to identify the presence of possible periodic oscillations: the
most significant oscillations occur at those log-frequencies that
make the spectrum $P_N(H,q)$ maximum.

\item Change $q$ and $H$ and redo the analysis. In this
work, we scan $H \in (-1,1)$ with spacing $0.1$ and $q \in
(0,1)$ with spacing $0.05$.

\item Having identified the strongest peak in the spectrum $P_N(H,q)$ and its
associated log-frequency, we study their dependence as
a function of $H$ and $q$.
\end{enumerate}

\begin{figure}
\begin{center}
\epsfig{file=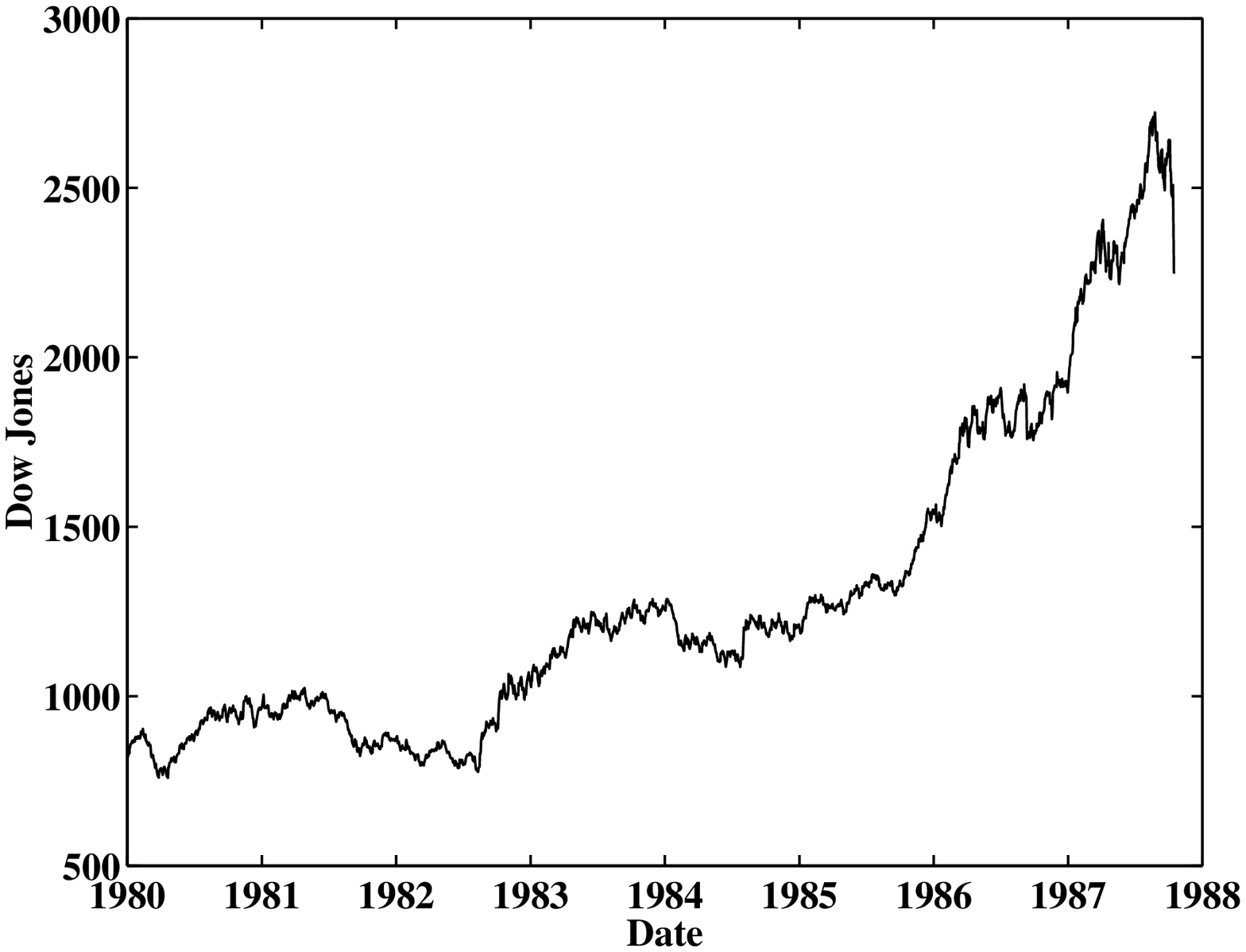,width=12cm, height=9cm}
\end{center}
\caption{The daily evolution of the Dow Jones Industrial Average
from 02-Jan-1980 to the ``Black Monday'' on 19-Oct-1987.}
\label{Fig:Dow87}
\end{figure}

Let us illustrate our procedure on the daily evolution of the Dow
Jones Industrial Average from 02-Jan-1980 to the ``Black Monday''
on 19-Oct-1987, shown in Fig.~\ref{Fig:Dow87}.

\begin{figure}
\begin{center}
\epsfig{file=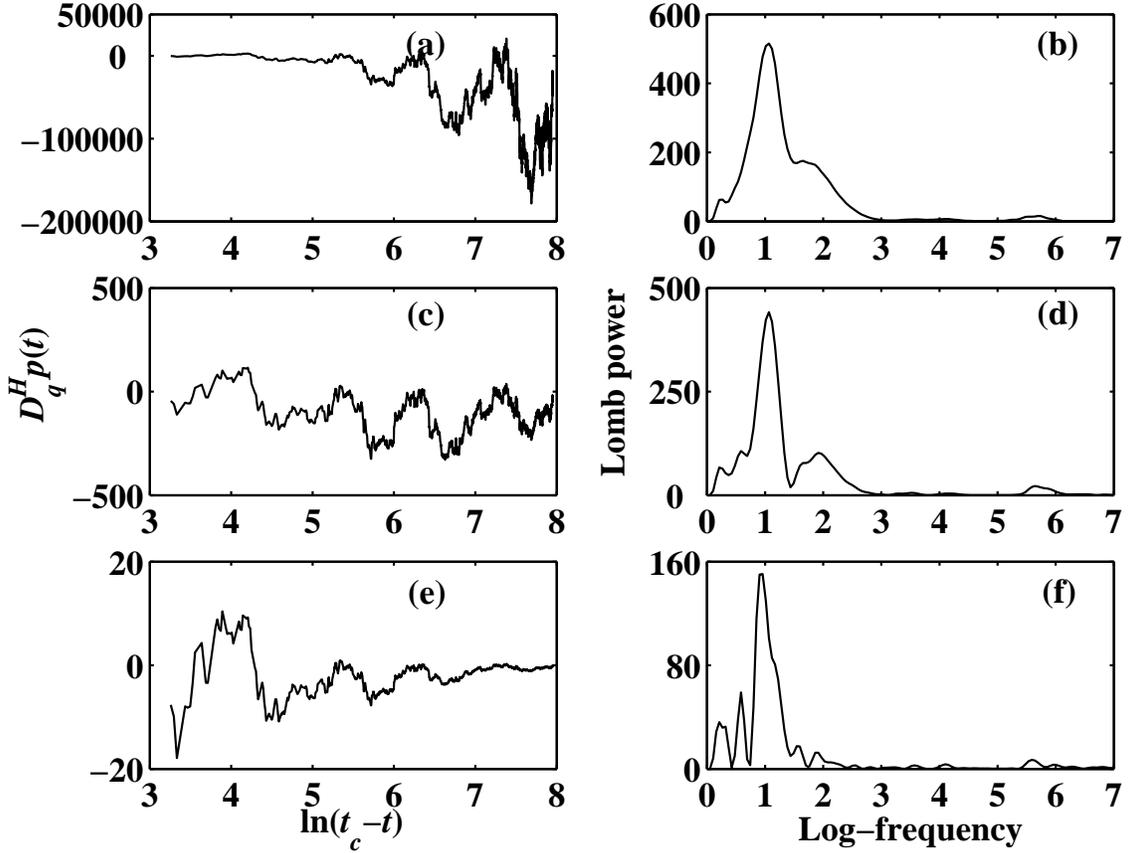, width=15cm, height=12cm}
\end{center}
\caption{Time dependence in the variable $\ln(t_c-t)$ of the
$(H,q)$-derivative (left
panels) for the Dow Jones time series shown in figure \ref{Fig:Dow87}
and their Lomb spectra
(right panels) for fixed $q = 0.65$ and varying $H$: (a-b) $H =
-0.9$; (c-d) $H =
0.1$; and (e-f) $H = 0.9$. The log-periodic structures of
$D_q^Hp(t)$ in plots (a), (c) and (e) is visible to the naked eye. The
amplitude of the log-periodic oscillations decreases with
increasing $H$ for fixed $q$. We observe peaks at $f_1 =
1.06$ in (b), at $f_1 = 1.06$ in (d) and at $f_1 = 0.94$ in (f).
We also find harmonics $f_2 = 1.94$ in (d) and $f_2 = 1.90$ in
(f). Recall the definition (\ref{Eq:Sor},\ref{deff}) of the log-frequency.}
\label{Fig:Dow87DqHLombP1}
\end{figure}

\begin{figure}
\begin{center}
\epsfig{file=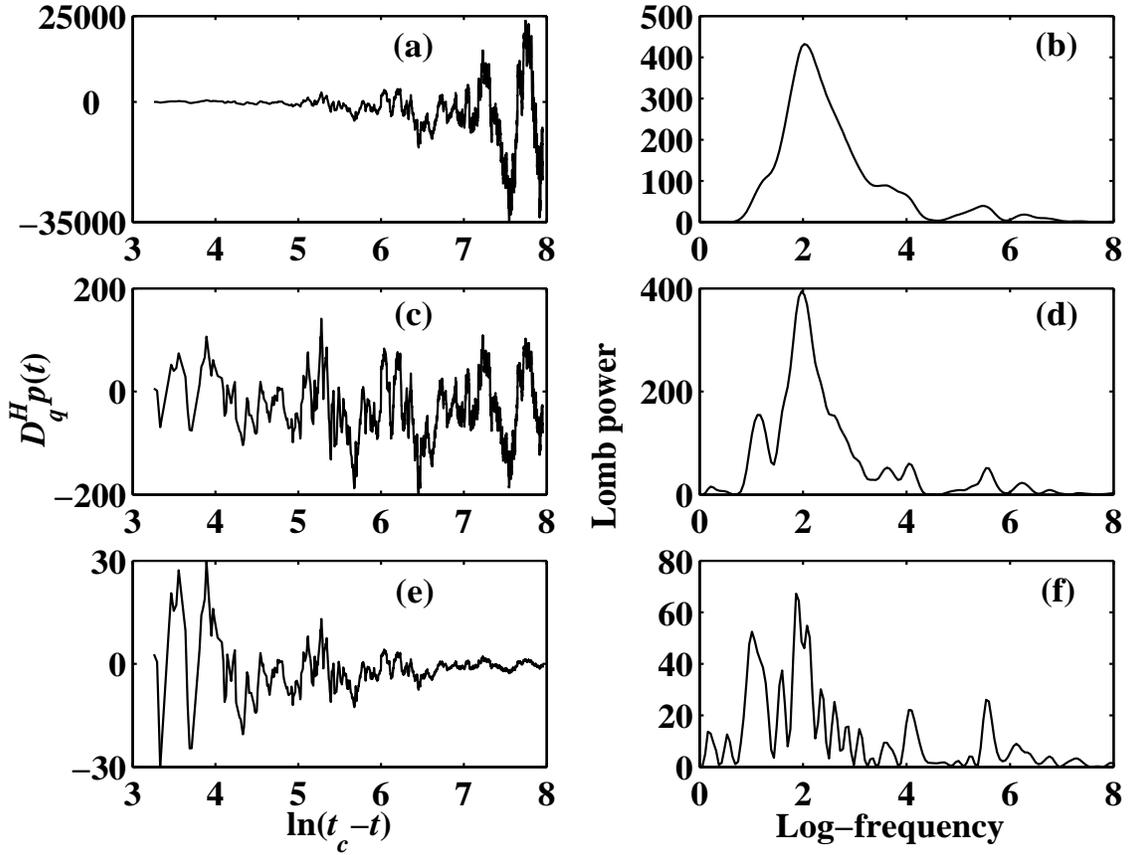, width=15cm, height=12cm}
\end{center}
\caption{Same as Fig.~\ref{Fig:Dow87DqHLombP1} but for $q = 0.9$:
(a-b) $H = -0.9$; (c-d) $H = 0.1$; and (e-f) $H = 0.9$.
The peaks of the Lomb spectra in plots (b) and (d) are very
high with log-frequencies $f_2=2.02$ and $1.96$ that can be interpreted as
harmonics $2f$ of the log-frequencies observed in
Fig.~\ref{Fig:Dow87DqHLombP1}.
We also see a peak at $f_1 = 1.15$ in (d) and peaks at $f_1
= 1.01$, $f_2 = 1.87$ and $f_4 = 4.06$ in (f), suggesting the observation
of several levels a discrete hierarchy.} \label{Fig:Dow87DqHLombP2}
\end{figure}

\begin{figure}
\begin{center}
\epsfig{file=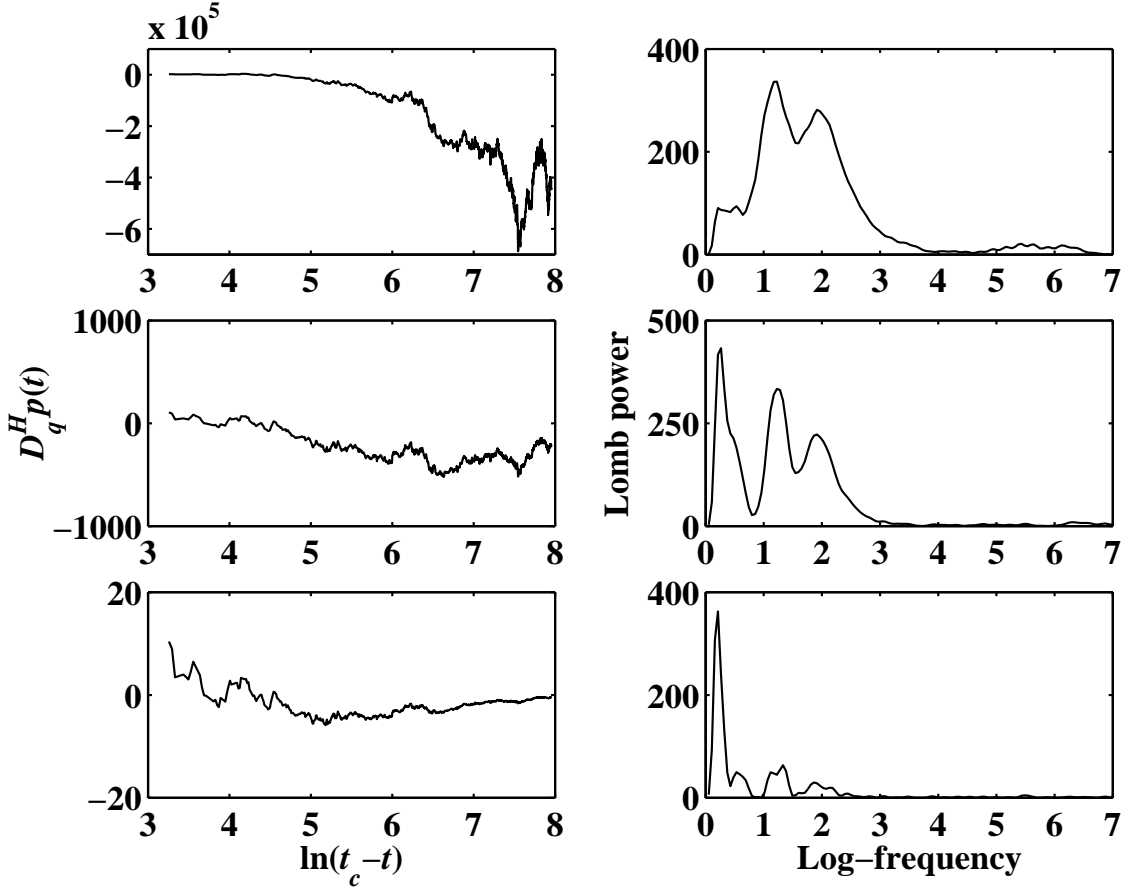, width=15cm, height=12cm}
\end{center}
\caption{Same as Fig.~\ref{Fig:Dow87DqHLombP1} but for $q = 0.3$.
The log-periodic structures of $D_q^Hp(t)$ in plots (a), (c) and
(e) are much weaker than in previous figures. Plot (b) illustrates that
the log-periodic structure in (a) is nevertheless very significant with
the log-frequencies $f_1 = 1.22$
and $f_2 = 1.92$ corresponding to the
highest peaks of the spectrum. In the plots (d) and (f),
we also observe secondary peaks at $f_1 = 1.22$ and $f_2 = 1.92$ in
(d) and $f_1 = 1.33$ and $f_2 = 1.86$ in (f) that are consistent with
previous values. However, the highest peak is in both cases found
for a very low log-frequency $f^{mp} =0.32$, that can be shown
to result from the trend induced by this choice of $q$. See text
for an explanation. }
\label{Fig:Dow87DqHLombB}
\end{figure}

In Figs.~\ref{Fig:Dow87DqHLombP1}, \ref{Fig:Dow87DqHLombP2} and
\ref{Fig:Dow87DqHLombB}, we present typical plots of the
generalized $q$-derivative and its Lomb periodogram for the Dow
Jones 1987 Crash. Figure \ref{Fig:Dow87DqHLombP1} shows the
evolution as a function of $\ln(t_c-t)$ of the generalized
$q$-derivative (left panel) and their Lomb periodograms (right
panel) for fixed $q = 0.65$: (a-b) $H = -0.9$; (c-d) $H = 0.1$;
and (e-f) $H = 0.9$. The log-periodic structures of $D_q^Hp(t)$ in
plots (a), (c) and (e) are clearly visible. The amplitude of the
log-periodic oscillations decreases with increasing $H$ for fixed
$q$. The envelops of the oscillations can be fitted well by the
expression
\begin{equation}
D_m = \left(B' \pm C'\sqrt{C_1^2+C_2^2}\right)(x_m)^{\beta -H}
\label{Eq:Dm}
\end{equation}
giving the local maxima of the log-periodic function
(\ref{Eq:Sor}). We observe a peak at $f_1 = 1.06$ in (b), $f_1 =
1.06$ in (d) and $f_1 = 0.94$ in (f). We also see harmonics $f_2 =
1.94$ in (d) and $f_2 = 1.90$ in (f). Figure
\ref{Fig:Dow87DqHLombP2} is for $q = 0.9$: (a-b) $H = -0.9$; (c-d)
$H = 0.1$; and (e-f) $H = 0.9$. The Lomb peaks in plots (b) and
(d) are very high with log-frequencies $f_2=2.02$ and $1.96$ that
may be interpreted as the second harmonics $2f$ of the frequency
found in figure \ref{Fig:Dow87DqHLombP1}. We find a peak at $f_1 =
1.15$ in (d) and peaks at $f_1 = 1.01$, $f_2 = 1.87$ and $f_4 =
4.06$, which makes more plausible the existence of harmonics of a
fundamental log-periodic component in the signal. Figure
\ref{Fig:Dow87DqHLombB} is for $q = 0.3$. The log-periodic
structures of $D_q^Hp(t)$ in plots (a), (c) and (e) are much less
obvious visually than in previous figures. Nevertheless, the Lomb
spectral analysis shown in panel (b) shows that the log-periodic
structure in (a) is significant at two log-frequencies $f_1 =
1.22$ and $f_2 = 1.92$. We note that the almost equal strength of
the two overlapping spectral peaks at these two frequencies is
probably causing a distortion in their estimation.

In panels (d) and (f) of figure \ref{Fig:Dow87DqHLombB},
the highest peaks occur at a significantly smaller log-frequency
which is found close to $0.32$. This log-frequency, which corresponds
to approximately
$1$ to $1.5$ full log-periodic oscillation in the data, can usually
be associated with
noise rather than a genuine signal. Indeed,
log-periodicity appears generically when
dealing with power laws in the presence of some kind of cumulative noise
\cite{fmp} or with noise
with long-range correlations \cite{Zhou02a}. The most probable
log-frequency resulting from noise is given by \cite{fmp}
\be
f^{mp} = {1.5 \over
\ln(t_c-t_{\min})-\ln(t_c-t_{\max})}~,
\label{Eq:fmp}
\ee
where
$t_{\min}$ and $t_{\max}$ are respectively the times defining the beginning and
the end of the time series. The value $0.32$ of the highest peaks in panels
(d) and (f) is compatible with $f^{mp}$ given by (\ref{Eq:fmp}).
We should therefore exclude from our statistics those pairs $(H,q)$
whose spectra are dominated
by the spurious low log-frequency $f^{mp}$ resulting from noise.

In addition, we should also discard from our statistics the pairs $(H,q)$
whose log-frequencies $f$ of the largest spectral peak
are very large, because a large log-frequency corresponds to many oscillations
which most often can be associated with high-frequency noise.
In Ref.~\cite{CriRup}, log-periodic components with
$2\pi f \geq 14$ were discarded for this reason.

Finally, let us note that panels (d) and (f)
of figure \ref{Fig:Dow87DqHLombB} also contains secondary peaks at
$f_1 = 1.22$ and $f_2 = 1.92$ in (d) and $f_1 = 1.33$ and $f_2 =
1.86$ in (f) which are not far from the log-frequencies previously
estimated.

\section{Log-periodicity of Wall Street Crashes}
\label{s:WallStreet}

We now exploit the procedure described in section \ref{ss:Criteria}
to analyze seven financial time series ending at
\begin{enumerate}
\item the Dow Jones Industrial Average, October 1987 Crash,
\item the Dow Jones Industrial Average, October 1997 strong correction,
\item the S\&P 500 Index, October 1987 crash,
\item the S\&P 500 Index, October 1997 strong correction,
\item the Nasdaq Index, October 1987 crash,
\item the Nasdaq Index, October 1997 strong correction, and
\item the Nasdaq Index, April 2000 crash.
\end{enumerate}

  In the spirit of
R. Feynman  \footnote{R. Feynman,
``Surely you're joking, Mr. Feynman'', Norton, 1985, page 341:
``If you're doing an experiment, you should report everything that you
think might make it invalid -- not only what you think is right about
it: other causes that could possibly explain your results; ... Details
that could throw doubt on your interpretation must be given, if you
know them. You must do the best you can --- if you know anything at
all wrong, or possibly wrong --- to explain it. If you make a theory,
for example, and advertise it, or put it out, then you must also put
down all the facts that disagree with it, as well as those that agree
with it. ... In summary, the idea is try to give ALL of the
information to help others to judge the value of your contribution;
not just the information that leads to judgment in one particular
direction or another.''}, we now describe extensively the evidence we
have accumulated on these seven time series, in particular by
presenting all the evidence that we think relevant, so that the reader
can judge for him- or her-self about the validity and robustness
of the results.

\subsection{Time series ending at the Dow Jones October 1987 Crash}

\begin{figure}
\begin{center}
\epsfig{file=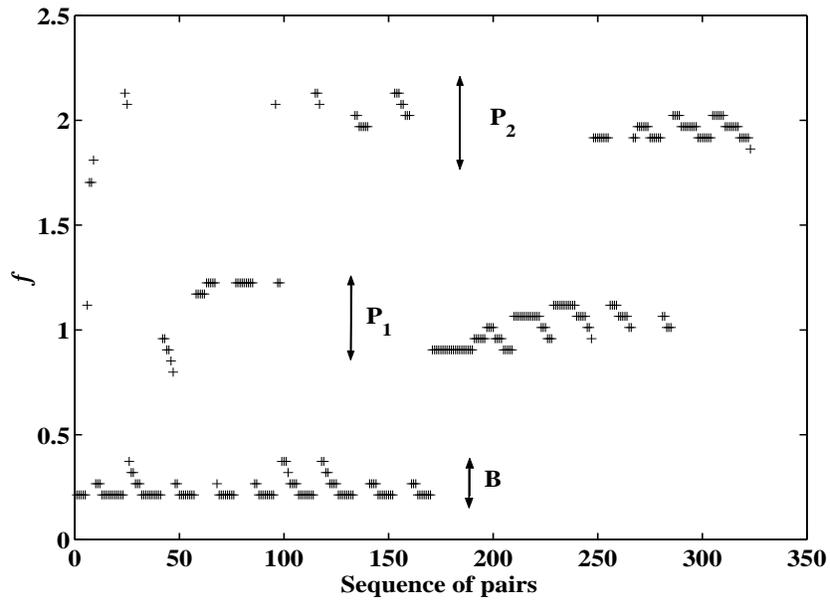, width=11cm, height=8cm}
\end{center}
\caption{For the Dow Jones time series shown in figure \ref{Fig:Dow87},
we report all the log-frequencies
corresponding to the highest peak of the Lomb spectrum obtained
for each of the pair $(H,q)$ that have been systematically sampled
with $H \in ]-1,1[$ with spacing $0.1$ and with $q \in
]0,1[$ with spacing $0.05$, providing a total of $18 \times 18$
``best'' log-frequencies.
The abscissa is counting the 324 pairs with an arbitrary choice
of indexing. Three clusters are clearly delineated:
cluster ${\mathbf{P_1}}$ with $f_1 = 1.04 \pm 0.11$, cluster
${\mathbf{P_2}}$ with
$f_2 = 1.98 \pm 0.07$ and cluster ${\mathbf{B}}$ with a log-frequency close
to $0.3$.}
\label{Fig:Dow87Pat}
\end{figure}

The log-periodic patterns characterizing the bubble phase
preceding this crash has been studied parametrically in
\cite{SJB,Vande,CriCrash99,CriCrash00,JohSorLed99}. We synthesize
the information obtained by our new non-parametric analyses such
as those shown in Figs.~\ref{Fig:Dow87DqHLombP1},
\ref{Fig:Dow87DqHLombP2} and \ref{Fig:Dow87DqHLombB} for all pairs
$(H, q)$ by plotting in figure \ref{Fig:Dow87Pat} the
log-frequency $f$ of the largest spectral peak as a function of an
index counting the pairs $(H, q)$. Since we scan $H \in (-1,1)$
with spacing $0.1$ and $q \in (0,1)$ with spacing $0.05$, this
corresponds to 324 pairs.
\begin{itemize}
\item cluster ${\mathbf{P_1}}$ centered on $f_1 = 1.04
\pm 0.11$ seems to be associated with a fundamental log-frequency in the data;
\item cluster ${\mathbf{P_2}}$ centered on $f_2 = 1.98 \pm 0.07$ can be
interpreted as the frequency $2 f_1$ of the second harmonics to the
fundamental frequency $f_1$;
\item cluster ${\mathbf{B}}$ centered on $f_{mp} \approx 0.3$ can be
attributed to the spurious log-frequency defined in (\ref{Eq:fmp}).
\end{itemize}

\begin{figure}
\begin{center}
\epsfig{file=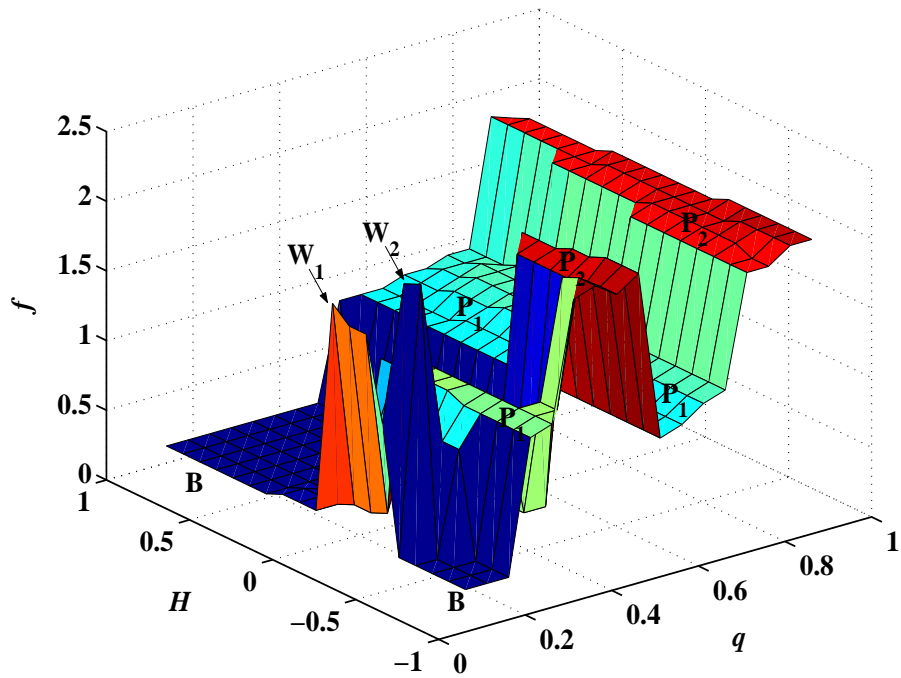, width=12cm, height=9cm}
\end{center}
\caption{Dependence of the log-frequency $f(H,q)$ of the
most significant peak in each Lomb spectrum of the
$(H,q)$-derivative for the  Dow Jones financial time series
ending at the October 1987 crash. Two
platforms ${\mathbf{P_1}}$ and ${\mathbf{P_2}}$ with $f_1 = 1.04
\pm 0.11$ and $f_2 = 1.98 \pm 0.07$ can be observed.} \label{Fig:Dow87f}
\end{figure}

Fig.~\ref{Fig:Dow87f} presents a more detailed description of the
log-frequency $f(H,q)$ of the most significant peak in each Lomb
periodogram of the $(H,q)$-derivative for the Dow Jones 1987
crash, by plotting $f(H,q)$ as a function of $H$ and $q$. The
clusters of figure \ref{Fig:Dow87Pat} now corresponds to
``platforms''. It is convenient to classify all pairs of $(H,q)$
into three classes based on the geometric shape of $f(H,q)$:
${\mathbf{W}}$ (wedge or wall), ${\mathbf{P}}$ (platform) and
${\mathbf{B}}$ (bottom of valley or basin). In figure
\ref{Fig:Dow87f}, we can clearly distinguish platforms
${\mathbf{P_1}}$ and ${\mathbf{P_2}}$ centered on $f_1 = 1.04 \pm
0.11$ and $f_2 = 1.98 \pm 0.07$. Two wedges are also indicated in
Fig.~\ref{Fig:Dow87f} as ${\mathbf{W_1}}$ and ${\mathbf{W_2}}$ but
their log-frequencies are not distinguishable from those of
${\mathbf{P_2}}$. The spurious cluster ${\mathbf{B}}$ is also
clearly visible.

\begin{figure}
\begin{center}
\epsfig{file=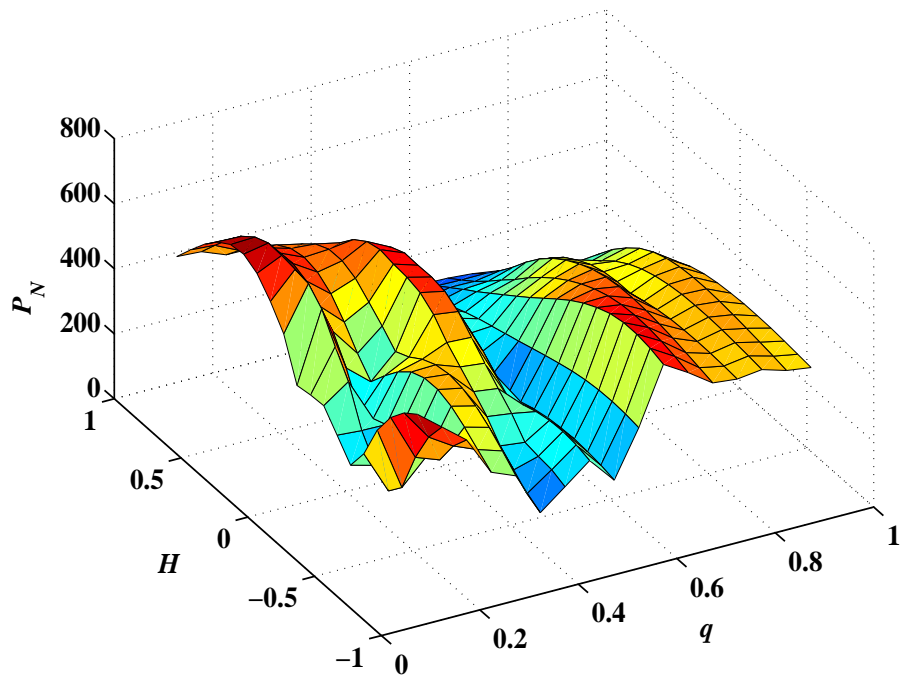, width=12cm, height=9cm}
\end{center}
\caption{Dependence of the highest peak $P_N(H,q)$ in each Lomb
spectrum of the $(H,q)$-derivative for the Dow Jones time series
ending at the October 1987 crash.} \label{Fig:Dow87PN}
\end{figure}

In order to compare the relative values of different $(H,q)$ pairs,
It is also instructive to construct in Fig.~\ref{Fig:Dow87PN} the
spectral height
$P_N(H,q)$ of the highest peak of each Lomb periodogram for each
$(H,q)$-derivative as a function of $H$ and $q$.
By matching Fig.~\ref{Fig:Dow87f} with
Fig.~\ref{Fig:Dow87PN}, we verify that the two parts of
${\mathbf{B}}$ have quite high Lomb peaks, while ${\mathbf{W_1}}$
and ${\mathbf{W_2}}$ in between correspond to valleys in
Fig.~\ref{Fig:Dow87PN}. In Fig.~\ref{Fig:Dow87f}, pairs $(H,q)$ with $q >
0.5$ lead to either $f_1$ or to its harmonic $f_2$. Small values of
$q$ thus give a weaker signal for log-periodicity. This property
will be seen to apply in all other time series discussed below.
We also note that on average, for
pairs $(H,q)$ with $q>0.5$, the largest Lomb spectral
peak increases with decreasing $H$. This is also illustrated in
Figs.~\ref{Fig:Dow87DqHLombP1}, \ref{Fig:Dow87DqHLombP2} and
\ref{Fig:Dow87DqHLombB}.

\begin{figure}
\begin{center}
\epsfig{file=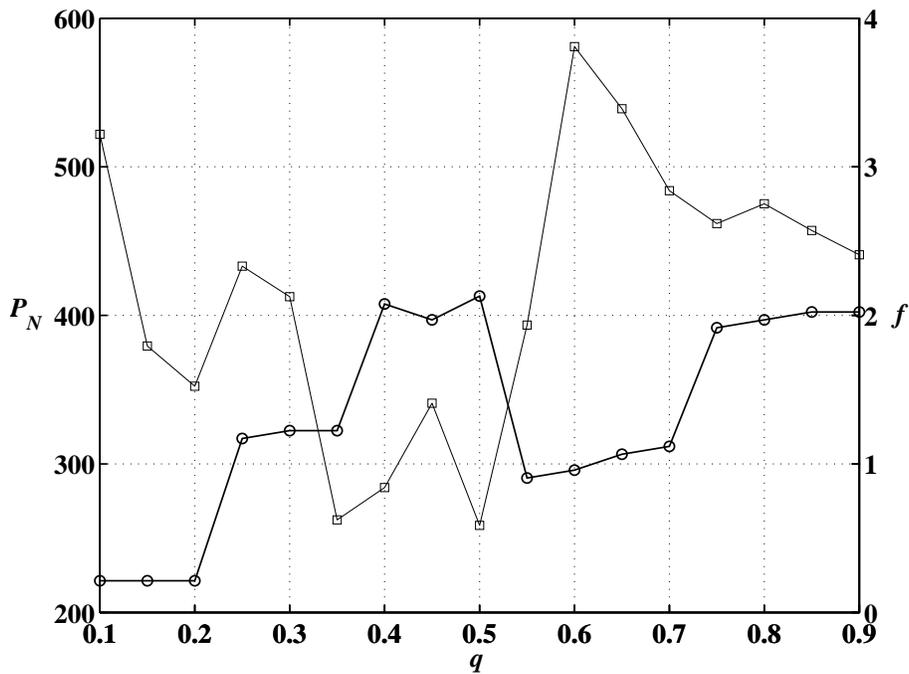, width=12cm, height=9cm}
\end{center}
\caption{Right vertical axis and thin line
marked with squares: dependence as a function of $q$ for fixed
$H=-0.7$ of the highest peak
$P_N(\hat{H},q)$ of the $(H,q)$-derivative of the Dow Jones time series
ending at the October 1987 crash. Left vertical axis and thick line
marked with circles: dependence as a function of $q$ for fixed $H=-0.7$ of
the log-frequency associated with these highest spectral peaks.
}
\label{Fig:Dow87PNf}
\end{figure}

Define an optimal pair $(\hat{H}, \hat{q})$ satisfying that $P_N
(\hat{H}, \hat{q})$ is the maximum of $P_N(H,q)$. Figure
\ref{Fig:Dow87PNf} shows on the left vertical coordinate and with
the thin line with squares a sectional cut of
Fig.~\ref{Fig:Dow87PN} for $H=-0.7$, that is, plots the highest
peak $P_N(\hat{H},q)$ as a function of $q$. The right vertical
coordinate and the thick line marked with circles correspond to a
sectional cut of Fig.~\ref{Fig:Dow87f} for the same value $H=-0.7$
and gives the log-frequency of the highest peak as a function of
$q$. The strongest spectral power is found for $q=0.6$ for which
the log-frequency is $f_1$.

\subsection{Time series ending with the Dow Jones October 1997 strong
correction}

The log-periodic patterns decorating the price trajectory of the
bubble phase preceding this strong correction have been already
studied parametrically in \cite{Feigen2,Van2,bookcrash}.

\begin{figure}
\begin{center}
\epsfig{file=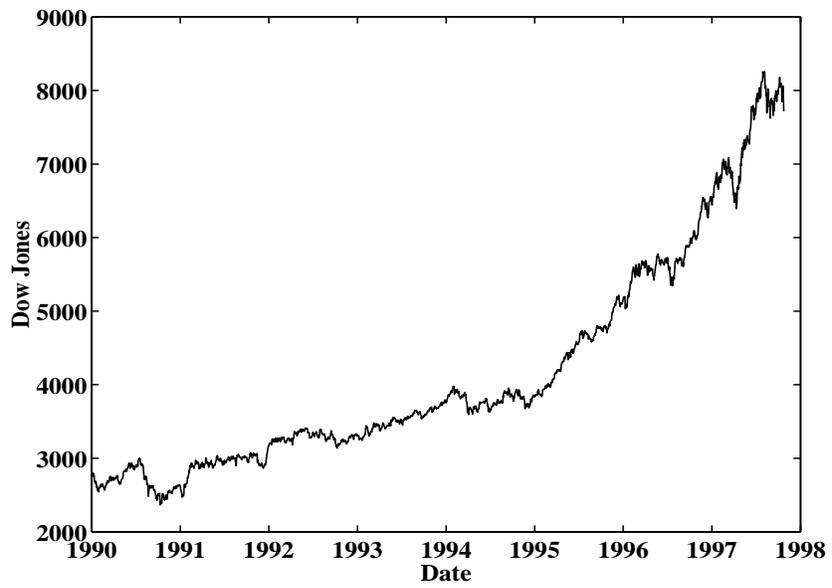,width=11cm, height=8cm}
\end{center}
\caption{The daily evolution of the Dow Jones Industrial Average
from 02-Jan-1990 to 27-Oct-1997.} \label{Fig:Dow97}
\end{figure}

\begin{figure}
\begin{center}
\epsfig{file=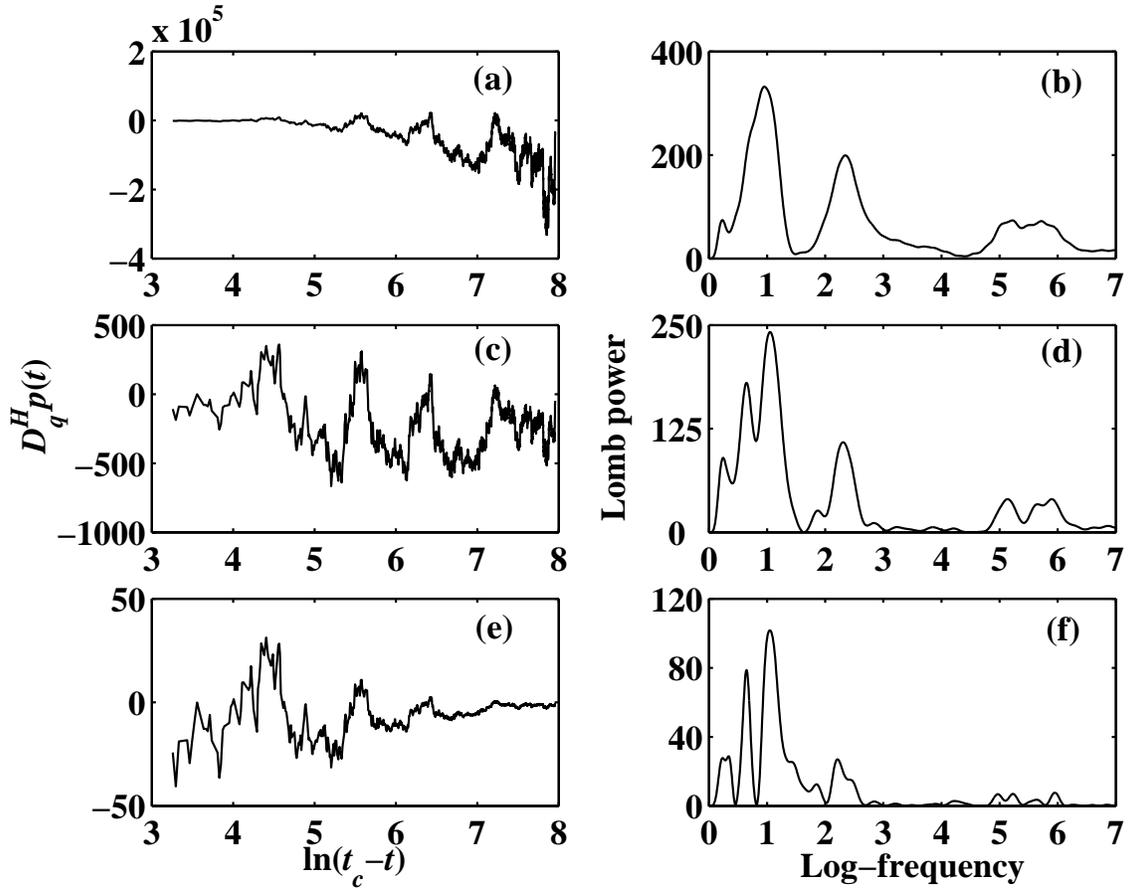, width=15cm, height=12cm}
\end{center}
\caption{The evolution of the generalized $q$-derivative (left
panels) for the Dow Jones time series ending at
the October 1997 strong correction and their Lomb spectra
(right panels) for fixed $q = 0.7$: (a-b) $H = -0.9$; (c-d) $H =
0.1$; and (e-f) $H = 0.9$. The highest peaks in the right panels
are: (b) $f_1 = 0.96$ and $f_2 = 2.34$; (d) $f_1 =
1.05$ and $f_2 = 2.31$; and (f) $f_1 = 1.04$ and $f_2 = 2.21$.}
\label{Fig:Dow97DqHLombP1}
\end{figure}

\begin{figure}
\begin{center}
\epsfig{file=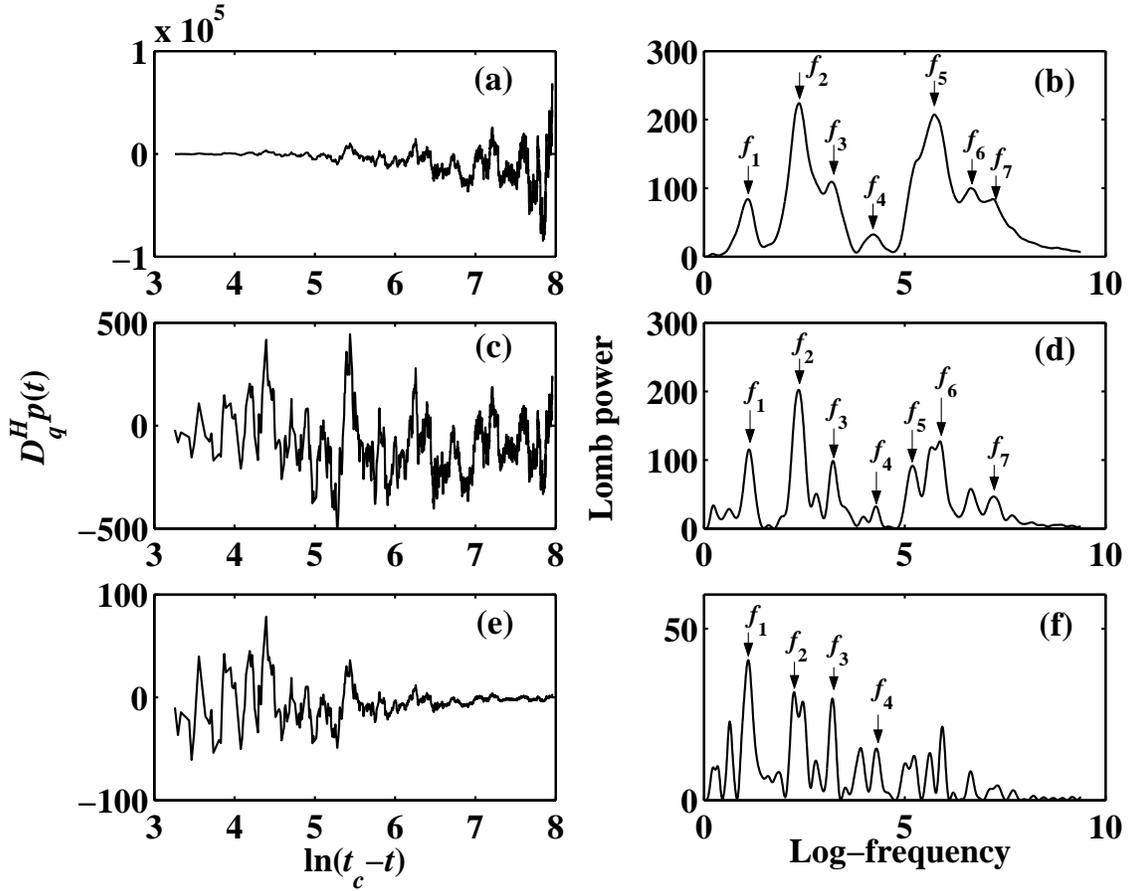, width=15cm, height=12cm}
\end{center}
\caption{Same as Fig.~\ref{Fig:Dow97DqHLombP1} for $q = 0.9$:
(a-b) $H = -0.9$; (c-d) $H = 0.1$; and (e-f) $H = 0.9$. The
log-frequencies and the associated harmonics are remarkably well-defined in the
Lomb spectra: (b) $f_1 = 1.10$, $f_2 = 2.36$, $f_3 = 3.18$,
$f_4 = 4.21$, $f_5 = 5.74$, $f_6 = 6.64$ and $f_7 = 7.21$; (d)
$f_1 = 1.12$, $f_2 = 2.35$, $f_3 = 3.21$, $f_4 = 4.28$, $f_5 =
5.19$, $f_6 = 5.87$ and $f_7 = 7.22$; and (f) $f_1 = 1.10$, $f_2 =
2.25$, $f_3 = 3.19$ and $f_4 = 3.91$. The amplitude of the
log-periodic oscillations decreases with increasing $H$ for fixed
$q$.} \label{Fig:Dow97DqHLombP2}
\end{figure}

\begin{figure}
\begin{center}
\epsfig{file=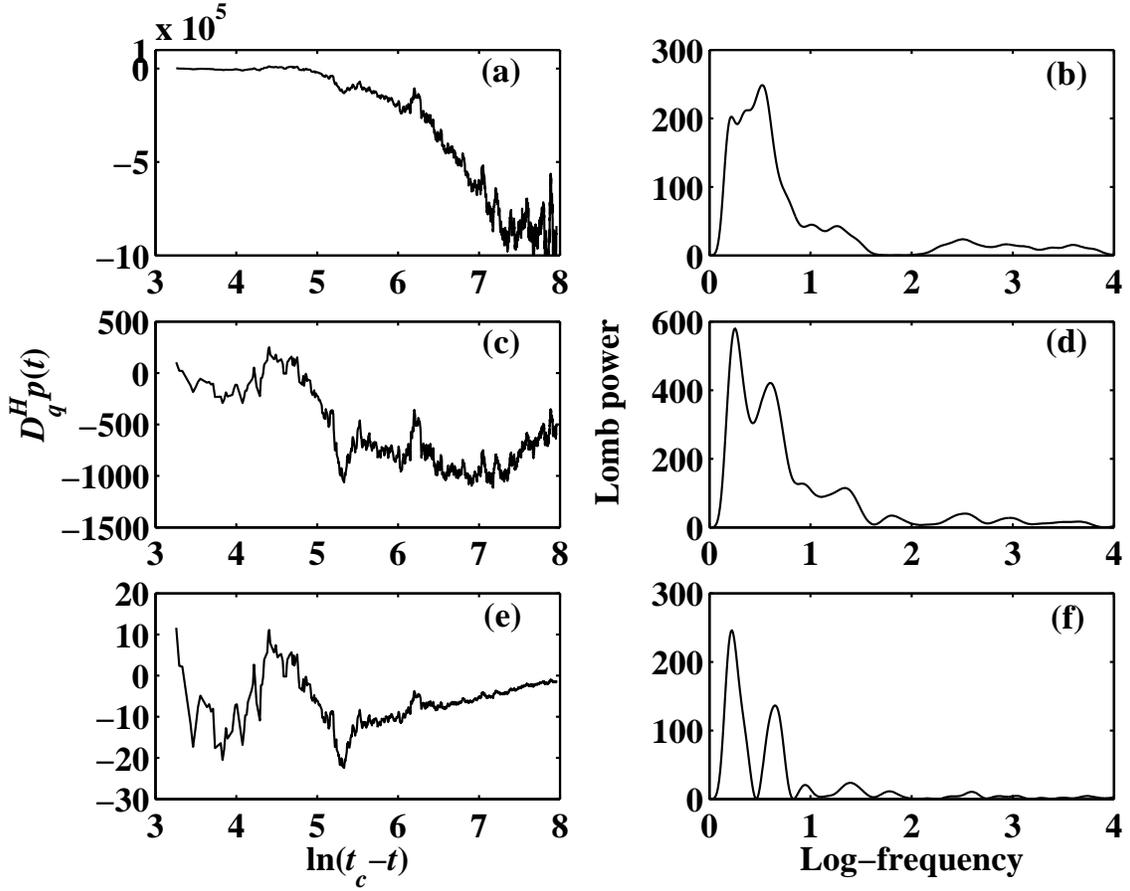, width=15cm, height=12cm}
\end{center}
\caption{Same as Fig.~\ref{Fig:Dow97DqHLombP1} for $q = 0.4$:
(a-b) $H = -0.9$; (c-d) $H = 0.1$; and (e-f) $H = 0.9$. The
log-periodic oscillations are ambiguous and the extracted
log-frequencies in (d) and (f) are close to the spurious
log-frequency $f^{mp}$. We can also see peaks at $f_{1/2} = 0.52$
in (b), $f_{1/2} = 0.61$ in (d) and $f_{1/2} = 0.65$ in (f), which
may be probably interpreted as sub-harmonics of the fundamental
log-frequency $f_1$.} \label{Fig:Dow97DqHLombP3B}
\end{figure}

\begin{figure}
\begin{center}
\epsfig{file=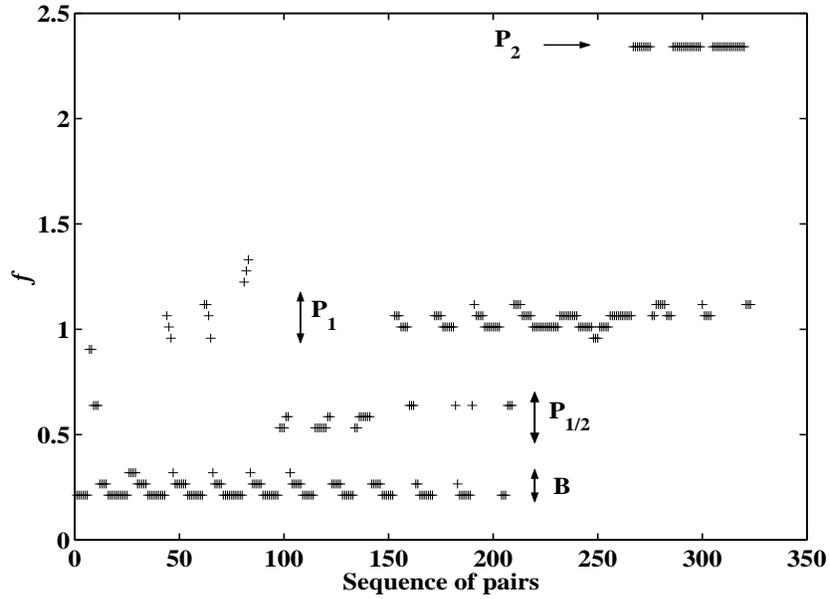, width=11cm, height=8cm}
\end{center}
\caption{Same as figure \ref{Fig:Dow87Pat} for the Dow Jones time
series ending at the October 1997 strong correction. Four clusters
${\mathbf{P_1}}$ with $f_1 = 1.04 \pm 0.05$, ${\mathbf{P_2}}$ with
$f_2 \equiv 2.34$, ${\mathbf{P_{1/2}}}$ with $f_{1/2} = 0.59 \pm
0.04$ and ${\mathbf{B}}$ are clearly visible. The small
log-frequency $f_B = 0.21$ of cluster ${\mathbf{B}}$ corresponds
to only one oscillation in the $(H,q)$-derivative.}
\label{Fig:Dow97Pat}
\end{figure}

\begin{figure}
\begin{center}
\epsfig{file=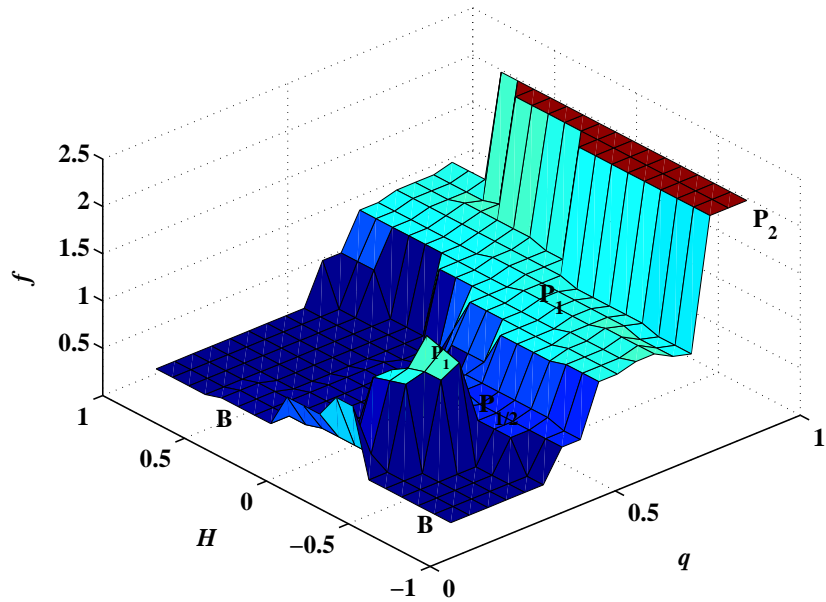, width=11cm, height=8cm}
\end{center}
\caption{Dependence of the log-frequency $f(H,q)$ of the
most significant peak in each Lomb periodogram of the
$(H,q)$-derivative for the
Dow Jones time series ending with the October 1997 strong correction.
The two platforms
${\mathbf{P_1}}$ and ${\mathbf{P_2}}$ have log-frequencies of $f_1
= 1.04 \pm 0.05$ and its $2f$ harmonic $f_2 \equiv 2.34$, while
${\mathbf{P_{1/2}}}$ shows probably what can be interpreted
as the subharmonic $f_{1/2} = 0.59 \pm 0.04$.} \label{Fig:Dow97f}
\end{figure}

\begin{figure}
\begin{center}
\epsfig{file=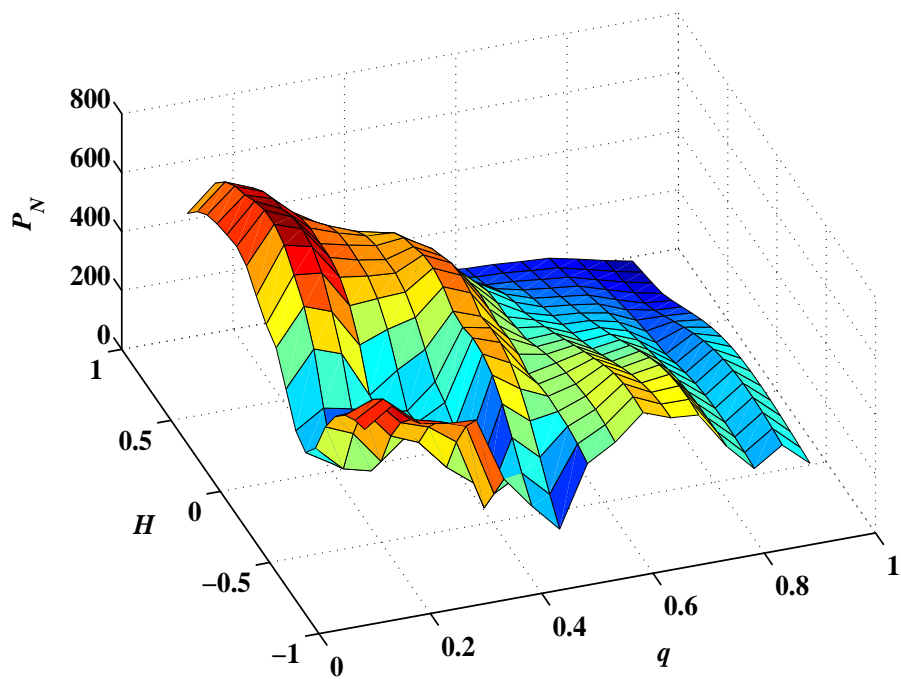, width=12cm, height=9cm}
\end{center}
\caption{Dependence of the highest peak $P_N(H,q)$ of each Lomb
periodogram of the $(H,q)$-derivative for the
Dow Jones time series ending with the October 1997 strong correction.}
\label{Fig:Dow97PN}
\end{figure}

\begin{figure}
\begin{center}
\epsfig{file=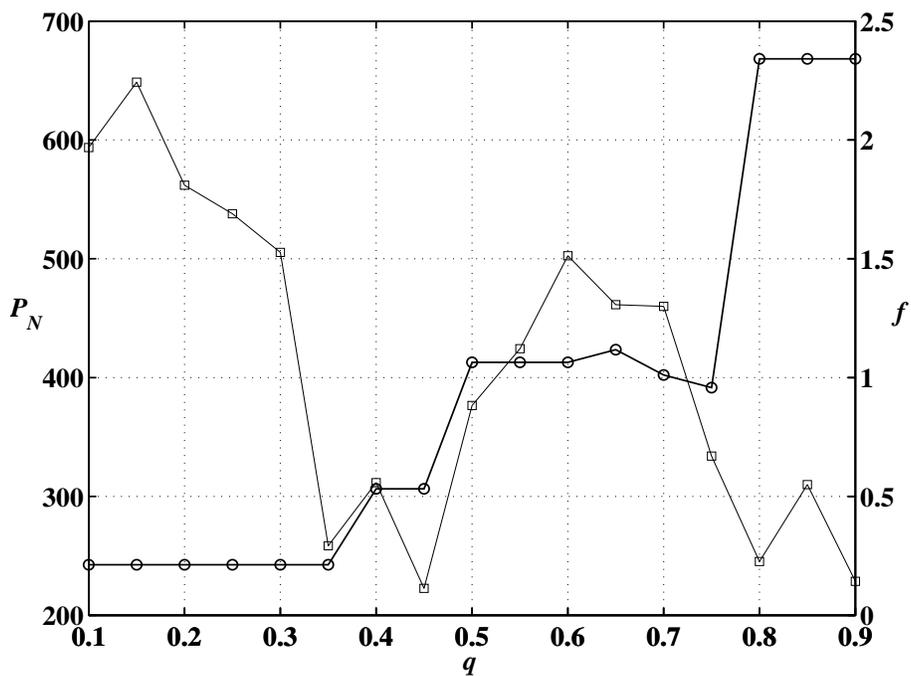, width=12cm, height=9cm}
\end{center}
\caption{Dependence as a function of $q$ for fixed
$H=-0.8$ of the highest spectral peak
$P_N(\hat{H},q)$ of the $(H,q)$-derivative, shown
on the left vertical axis with thin line
marked with squares and of the associated log-frequencies
$f(\hat{H},q)$ on the right vertical axis with the thick line
marked with circles, for the
Dow Jones time series ending with the October 1997 strong correction.}
\label{Fig:Dow97PNf}
\end{figure}

Figure \ref{Fig:Dow97} shows the daily evolution of the Dow Jones
Industrial Average from 02-Jan-19890 to 19-Oct-1997. On Monday,
September 27th, the Dow Jones index suffered from a strong loss of
more than $7\%$ that ended a strong bull regime of strongly
appreciating prices. Many observers believed on this day that it
was the starting point of a crash. The next day, a rally of $+5\%$
alleviated fears and started a market phase that last
approximately 3 months in which the market remained essentially
flat. The results of the analysis of the $(H,q)$-derivative of
this time series shown in figures \ref{Fig:Dow97DqHLombP1} to
\ref{Fig:Dow97PNf} are found to be very similar to those reported
previously for the Dow Jones time series of ending with the
October 1987 crash. Two novel features appear. First, figure
\ref{Fig:Dow97Pat} shows the existence of an additional cluster
${\mathbf{P_{1/2}}}$ at $f_{1/2} = 0.59 \pm 0.04$, that it is
tempting to interpret as the first sub-harmonic of the fundamental
log-frequency $f_1$. Second, for $q=0.9$, we observe in figure
\ref{Fig:Dow97DqHLombP2} several harmonics beyond the second
harmonics, suggesting a very strong log-periodicity with more
complex nonlinearity that for the October 1987 crash.

\subsection{Time series ending with the S\&P 500 Index, October 1987 crash}

\begin{figure}
\begin{center}
\epsfig{file=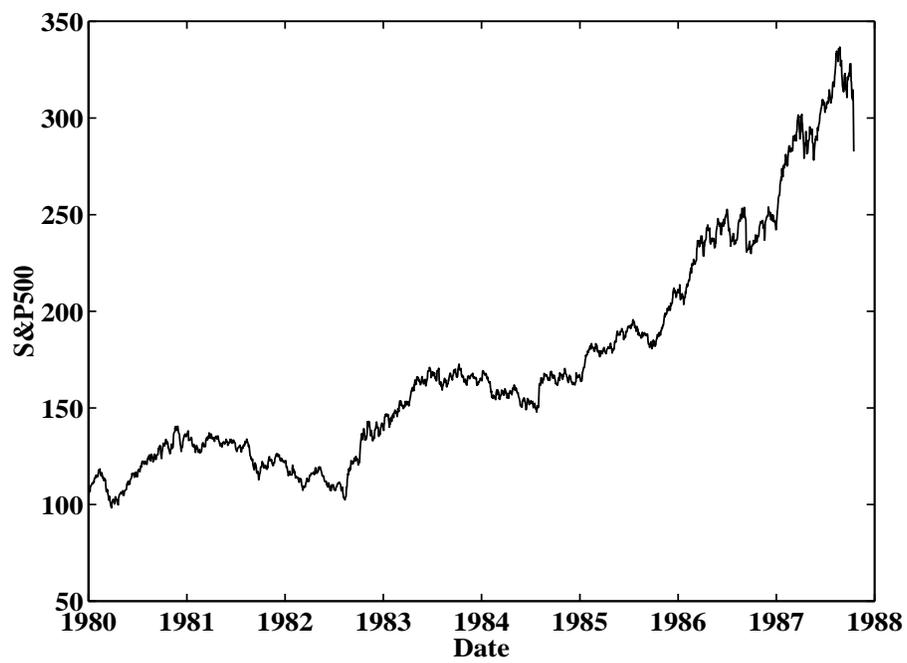,width=12cm, height=9cm}
\end{center}
\caption{Daily evolution of the S\&P 500 Index from 02-Jan-1980 to
the ``Black Monday'' on 19-Oct-1987.} \label{Fig:SP87}
\end{figure}

Figure \ref{Fig:SP87} shows the Daily evolution of the S\&P 500
Index from 02-Jan-1980 to the ``Black Monday'' on 19-Oct-1987,
which started the crash of October 1987. The time series is
obviously not independent of the Dow Jones time series shown in
figure \ref{Fig:Dow87} as the S\&P 500 Index contains the 30 blue
chip stocks constituting the Dow Jones index. However, it is
influenced by the behavior of 470 other stocks and we think it
useful to carry out a detailed analysis of this time series as
well, to compare with the Dow Jones time series and develop a
sense about how different realization of the noise and of the
signal may modify the log-periodicity.

\begin{figure}
\begin{center}
\epsfig{file=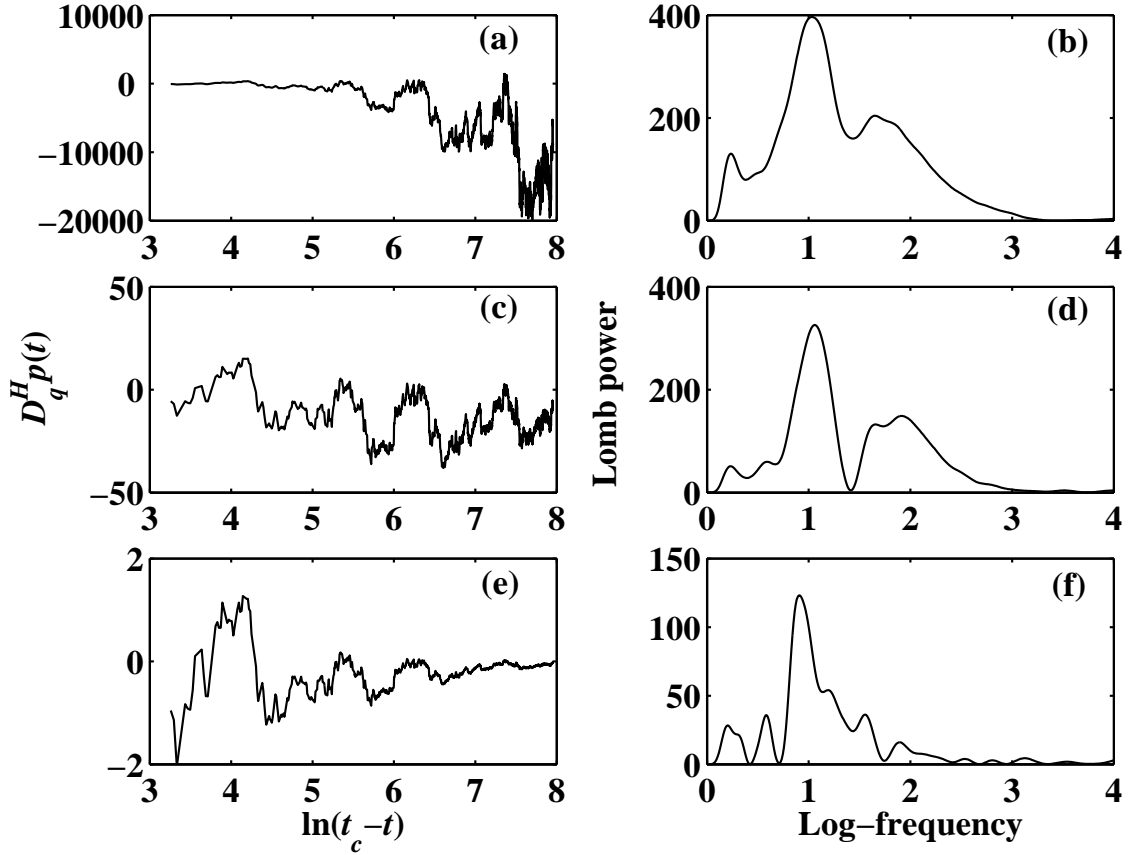, width=15cm, height=12cm}
\end{center}
\caption{Left panels: evolution of the generalized $q$-derivative
for the S\&P 500 time series ending with the October 1987 Crash
and their Lomb periodograms (right panel) for fixed $q = 0.6$:
(a-b) $H = -0.9$; (c-d) $H = 0.1$; and (e-f) $H = 0.9$. The
log-periodic structures of $D_q^Hp(t)$ in plots (a), (c) and (e)
are clearly visible. The amplitude of the log-periodic
oscillations decreases with increasing $H$ for fixed $q$. The
envelops of the oscillations can be fitted very well with
Eq.~(\ref{Eq:Dm}). We see Lomb peaks at $f_1 = 1.03$ in (b), $f_1
= 1.06$ in (d) and $f_1 = 0.91$ in (f) which are very significant.
We can also observe the harmonics  $f_2 = 1.65$ in (b) which is
reddened by the highest peak and $f_2 = 1.92$ in (d).}
\label{Fig:SP87DqHLombP1}
\end{figure}

\begin{figure}
\begin{center}
\epsfig{file=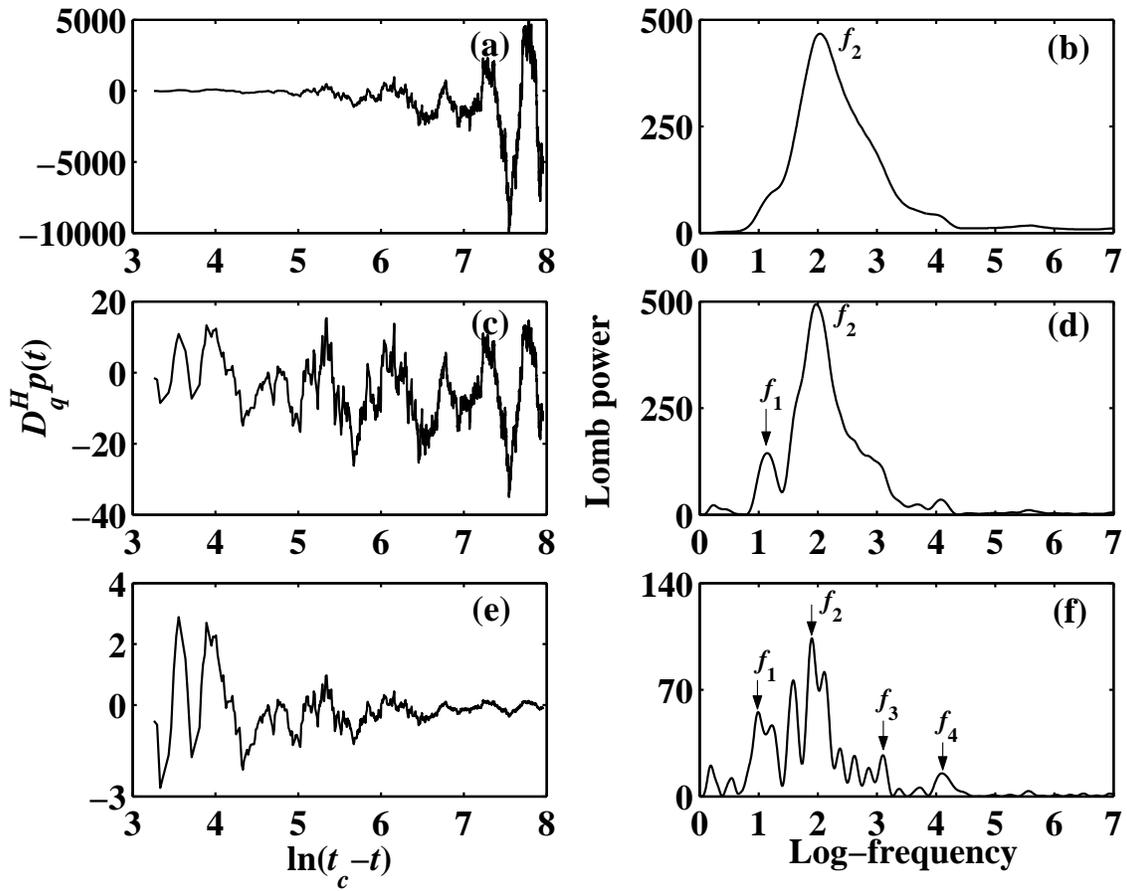, width=15cm, height=12cm}
\end{center}
\caption{Same as Fig.~\ref{Fig:SP87DqHLombP1} for $q = 0.85$:
(a-b) $H = -0.9$; (c-d) $H = 0.1$; and (e-f) $H = 0.9$.
The spectral peaks in panels (b) and (d) are very
high with log-frequencies $f_2=2.03$ and $1.97$ corresponding
to the second harmonics of $f_1$.
We can also see a peak at $f_1 = 1.14$ in (d) and peaks at $f_1
= 0.99$, $f_2 = 1.90$, $f_3 = 3.10$ and $f_4 = 4.10$ in (f) that
suggest a strong log-periodicity with its several harmonics.
}
\label{Fig:SP87DqHLombP2}
\end{figure}

\begin{figure}
\begin{center}
\epsfig{file=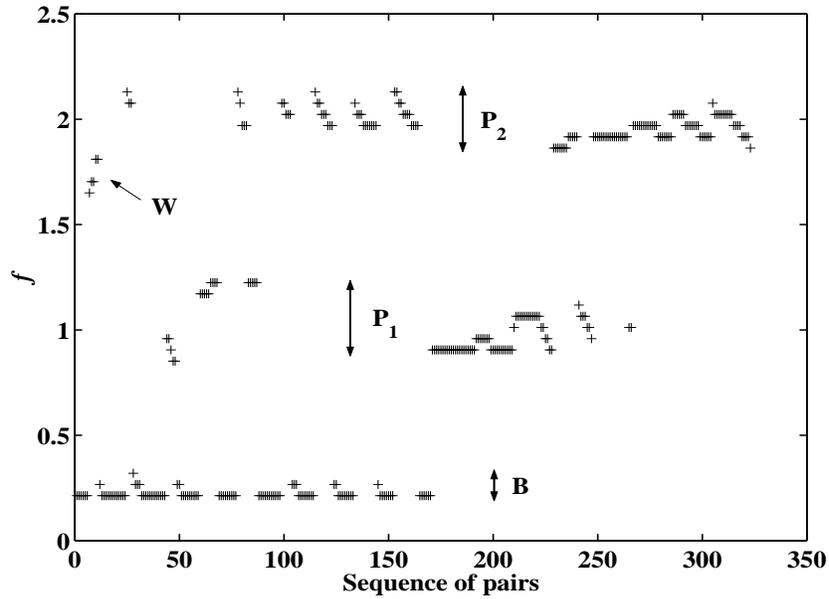, width=11cm, height=8cm}
\end{center}
\caption{Same as figure \ref{Fig:Dow87Pat} for the
S\&P 500 time series ending at
the October 1987 crash. Four clusters can be observed:
${\mathbf{P_1}}$ with $f_1 = 1.00 \pm 0.11$, ${\mathbf{P_2}}$ with
$f_2 = 1.97 \pm 0.06$ and ${\mathbf{B}}$ with $f_B = 0.21$ probably
associated with the spurious log-frequency $f_{mp}$; the fourth
``wedge'' cluster ${\mathbf{W}}$ has
log-frequency $f_{W} = 1.74 \pm 0.07$.}
\label{Fig:SP87Pat}
\end{figure}

\begin{figure}
\begin{center}
\epsfig{file=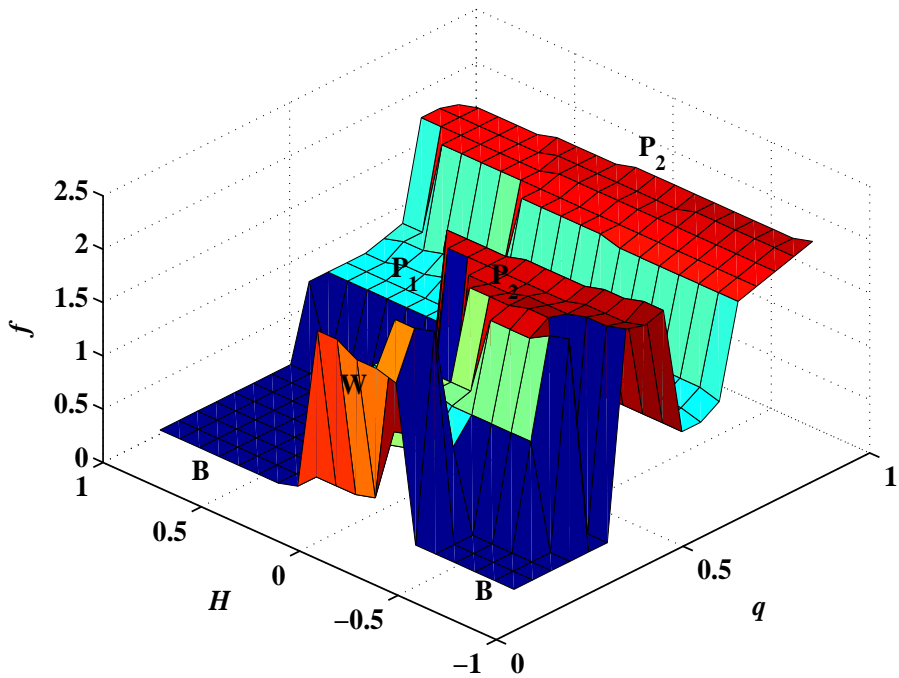, width=12cm, height=9cm}
\end{center}
\caption{Dependence of the log-frequency $f(H,q)$ of the most
significant peak in each Lomb spectrum of each $(H,q)$-derivative
for the S\&P 500 time series ending with the October 1987 crash.
We observe two platforms ${\mathbf{P_1}}$ and ${\mathbf{P_2}}$
with $f_1 = 1.00 \pm 0.11$ and $f_2 = 1.97 \pm 0.06$
respectively.} \label{Fig:SP87f}
\end{figure}

\begin{figure}
\begin{center}
\epsfig{file=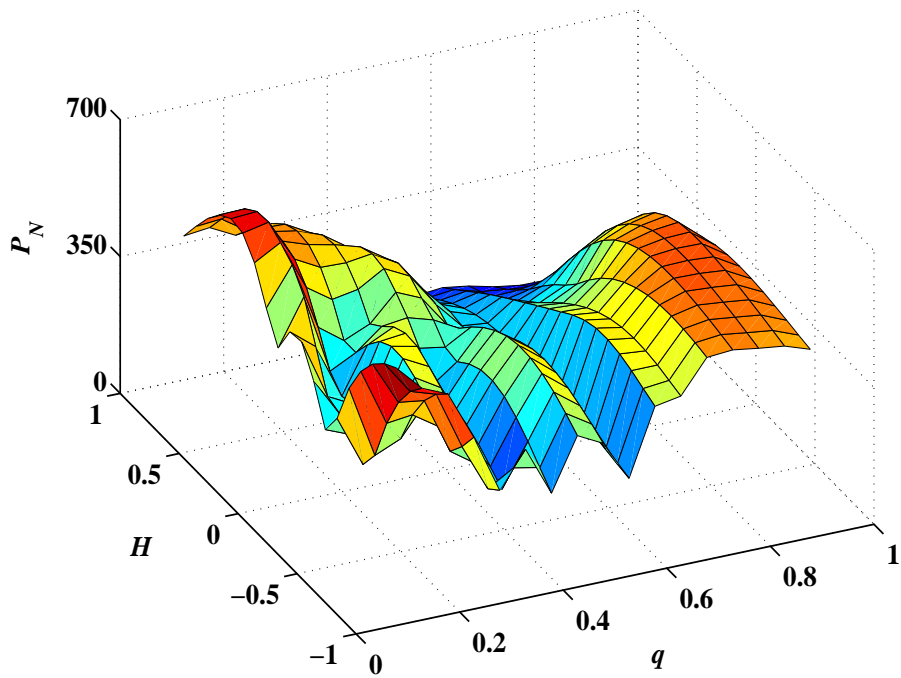, width=12cm, height=9cm}
\end{center}
\caption{Dependence of the highest peak $P_N(H,q)$ in each Lomb
spectrum of the $(H,q)$-derivative for the S\&P 500 time series
ending with the October 1987 crash.} \label{Fig:SP87PN}
\end{figure}

\begin{figure}
\begin{center}
\epsfig{file=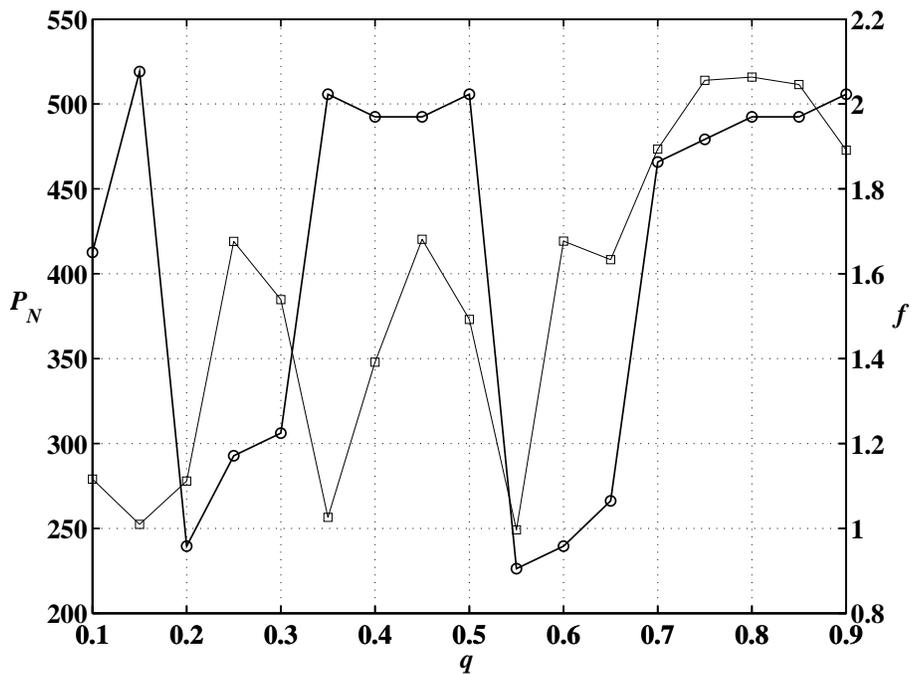, width=12cm, height=9cm}
\end{center}
\caption{Dependence as a function of $q$ for fixed $H=-0.3$ of the
highest peak $P_N(\hat{H},q)$ shown as the thin line marked with
squares with scales on the left vertical axis and of the
associated log-frequencies $f(\hat{H},q)$ shown as the thick line
marked with circles with scales on the right vertical axis, for
the S\&P 500 time series ending with the October 1987 crash.}
\label{Fig:SP87PNf}
\end{figure}

The results of the analysis of the $(H,q)$-derivative of this time
series shown in figures \ref{Fig:SP87DqHLombP1} to
\ref{Fig:SP87PNf} are found to be very similar to those reported
previously in the two previous time series. The same fundamental
log-frequency $f_1$ and its harmonics are observed. The only
really novel feature is the appearance of what is probably a
spurious ``wedge'' cluster  ${\mathbf{W}}$ with log-frequency
$f_{W} = 1.74 \pm 0.07$ shown in figure \ref{Fig:SP87Pat}.

\subsection{Time series ending with the S\&P 500 Index, October 1997
strong correction}

\begin{figure}
\begin{center}
\epsfig{file=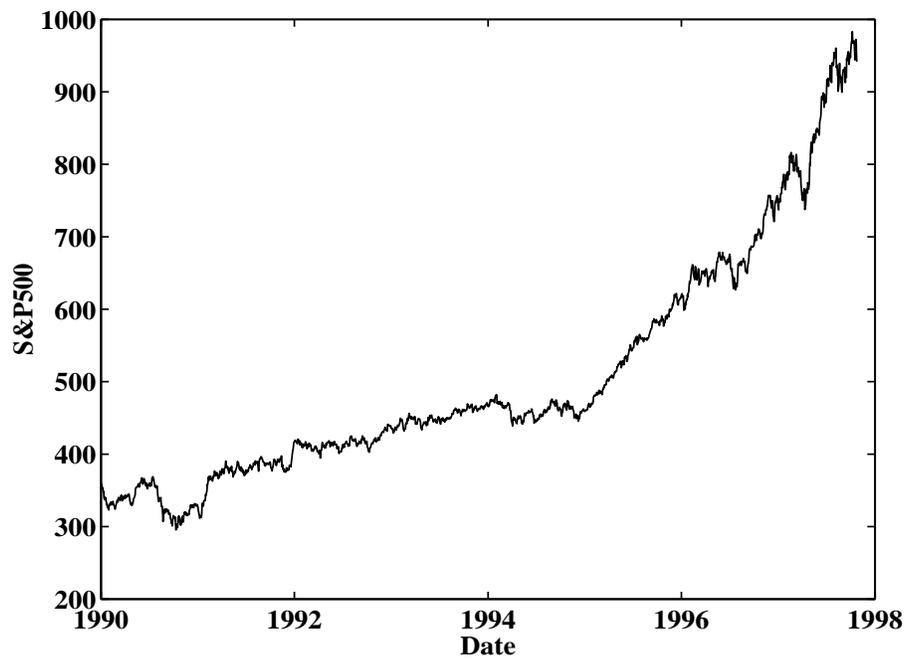,width=12cm, height=9cm}
\end{center}
\caption{Daily evolution of the S\&P 500 Index from 02-Jan-1990 to
27-Oct-1997.} \label{Fig:SP97}
\end{figure}

\begin{figure}
\begin{center}
\epsfig{file=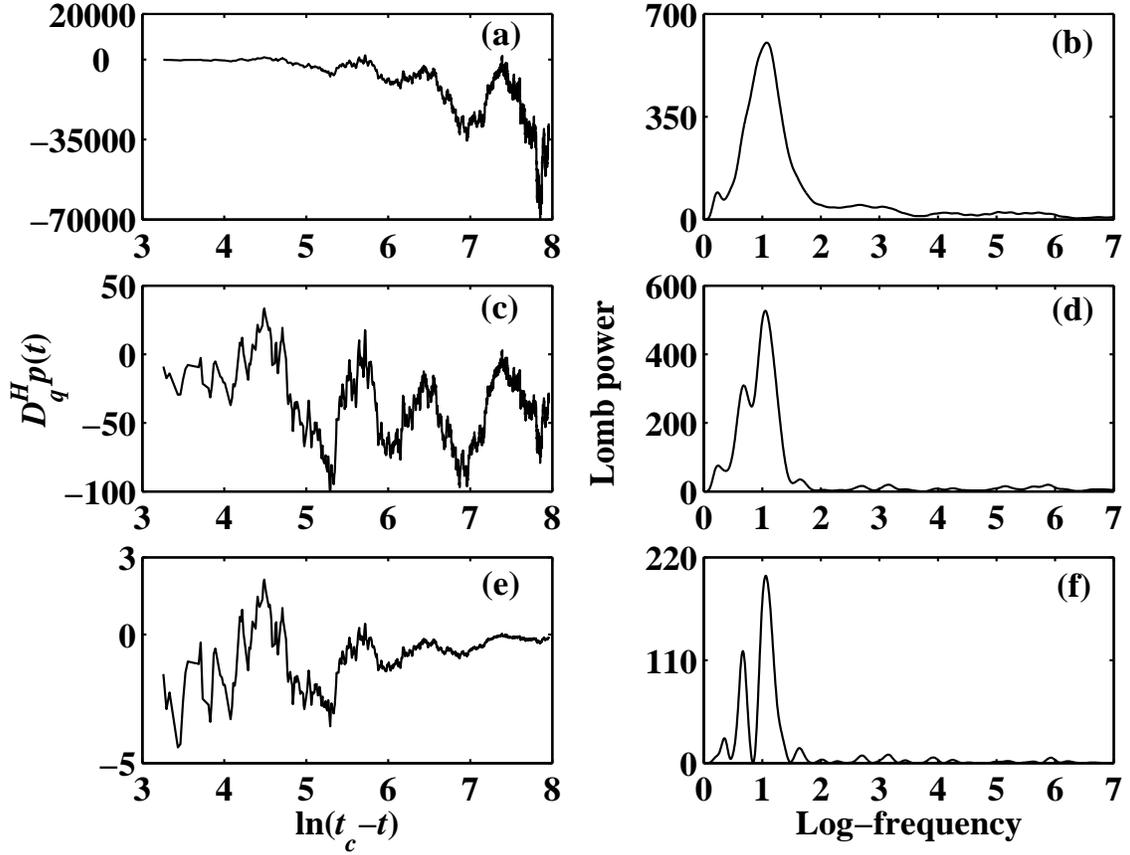, width=15cm, height=12cm}
\end{center}
\caption{Evolution of the $(H,q)$-derivative (left panel) for the
S\&P 500 time series ending with the 1997 strong correction and
their Lomb spectra (right panel) for fixed $q = 0.65$: (a-b) $H =
-0.9$; (c-d) $H = 0.1$; and (e-f) $H = 0.9$. The log-periodic
structures of $D_q^Hp(t)$ in plots (a), (c) and (e) are clearly
visible. The amplitude of the log-periodic oscillations decreases
with increasing $H$ for fixed $q$. The envelops of the
oscillations can be fitted very well with Eq.~(\ref{Eq:Dm}). We
see Lomb peaks at $f_1 = 1.07$ in (b), $f_1 = 1.06$ in (d) and
$f_1 = 1.06$ in (f), which are very significant. We can also
distinguish the sub-harmonics in the right panels.}
\label{Fig:SP97DqHLombP1}
\end{figure}

\begin{figure}
\begin{center}
\epsfig{file=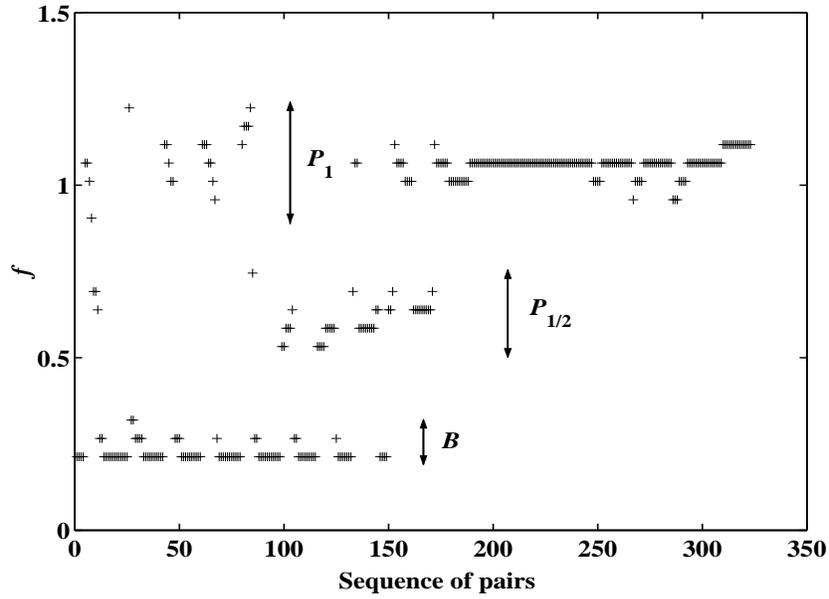, width=11cm, height=8cm}
\end{center}
\caption{Same as figure \ref{Fig:Dow87Pat} for the
S\&P 500 time series ending at
the October 1997 strong correction. Three clusters can be observed:
${\mathbf{P_1}}$ with $f_1 = 1.06 \pm 0.04$,
${\mathbf{P_{1/2}}}$ with $f_{1/2} = 0.61 \pm 0.05$ and ${\mathbf{B}}$
with $f_B = 0.21$.}
\label{Fig:SP97Pat}
\end{figure}

\begin{figure}
\begin{center}
\epsfig{file=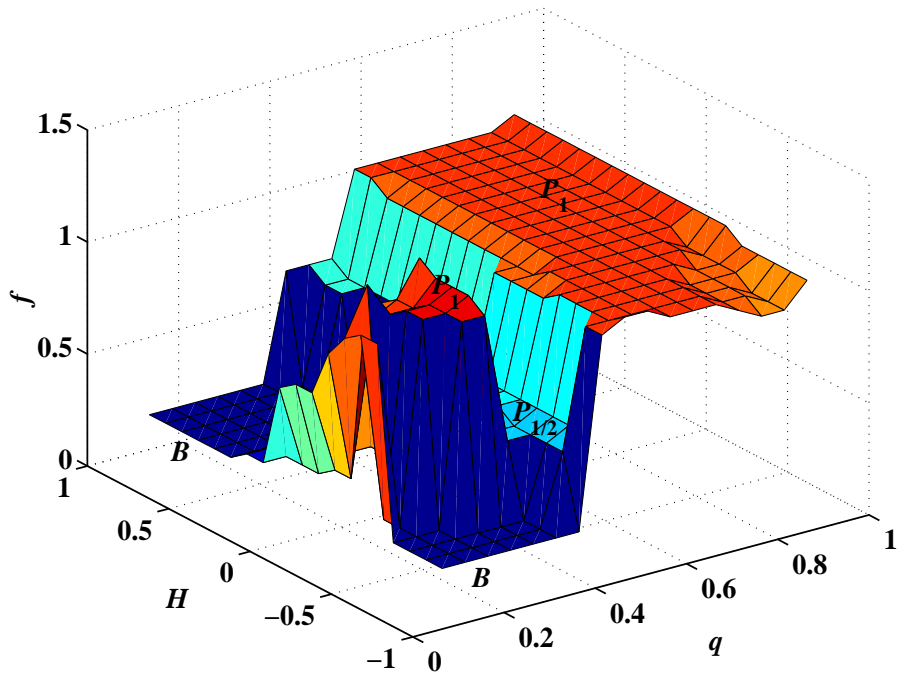, width=12cm, height=9cm}
\end{center}
\caption{Dependence of the log-frequency $f(H,q)$ of the most
significant peak in each Lomb periodogram of the
$(H,q)$-derivative for the S\&P 500 time series ending with the
October 1997 strong correction. We see clearly two platforms
${\mathbf{P_1}}$ and ${\mathbf{P_{1/2}}}$ with $f_1 = 1.06 \pm
0.04$ and $f_{1/2} = 0.61 \pm 0.05$.} \label{Fig:SP97f}
\end{figure}

\begin{figure}
\begin{center}
\epsfig{file=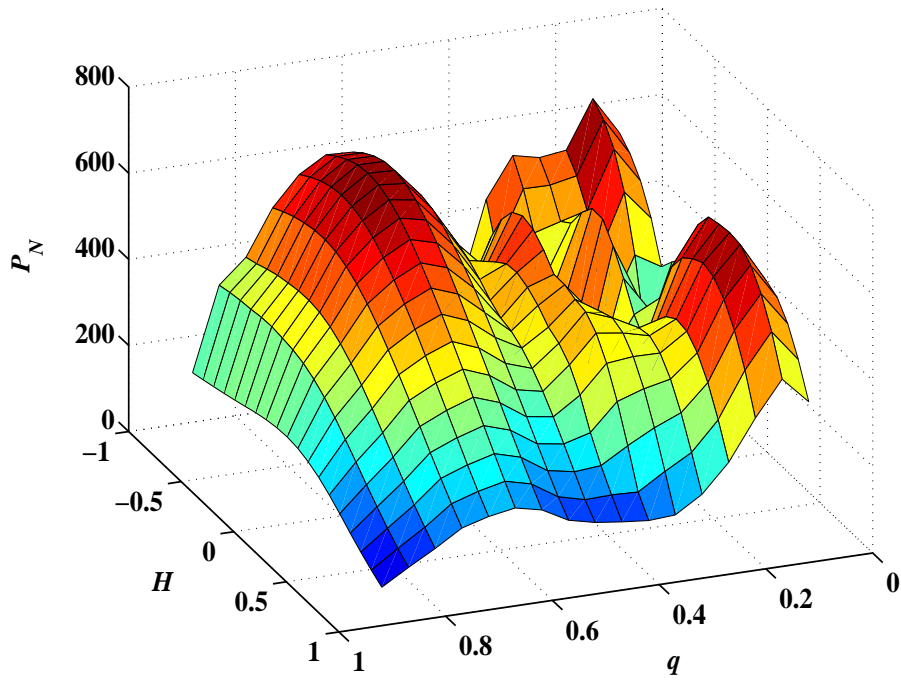, width=12cm, height=9cm}
\end{center}
\caption{Dependence of highest peak $P_N(H,q)$ in each Lomb
periodogram of the $(H,q)$-derivative for the S\&P 500 time series
ending with the October 1997 strong correction.}
\label{Fig:SP97PN}
\end{figure}

\begin{figure}
\begin{center}
\epsfig{file=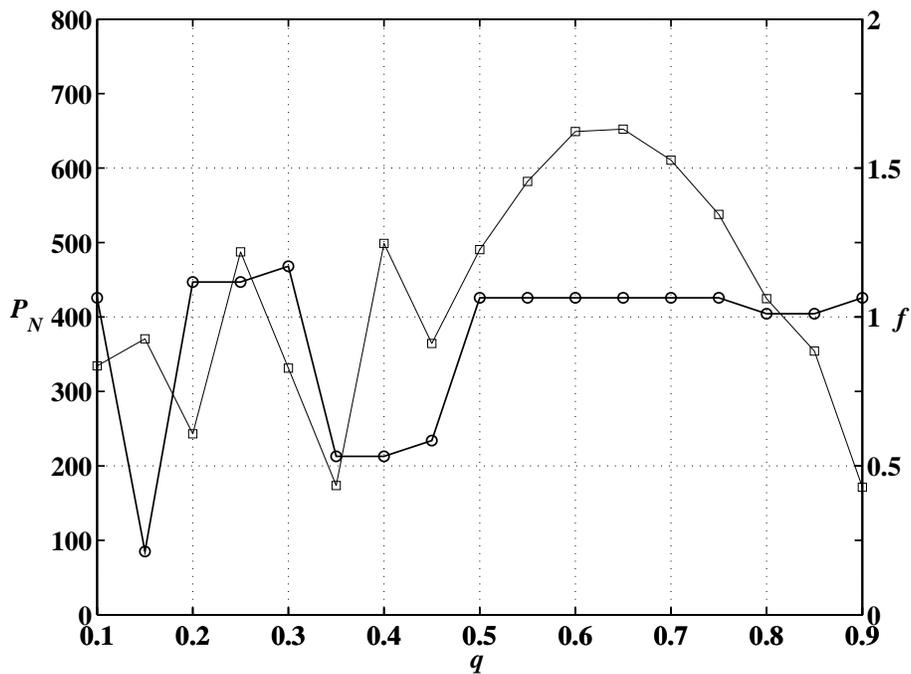, width=12cm, height=9cm}
\end{center}
\caption{Dependence with $q$ for $H=-0.5$ of the highest spectral
peak $P_N(\hat{H},q)$ shown as the thin line marked with squares
with left vertical scales and of the associated log-frequencies
$f(\hat{H},q)$ shown as the thick line marked with circles with
right vertical axis, for the S\&P 500 time series ending with the
October 1997 strong correction.} \label{Fig:SP97PNf}
\end{figure}

Figure \ref{Fig:SP97} shows the daily evolution of the S\&P 500
Index from 02-Jan-1990 to 27-Oct-1997. Similarly to the comparison
of the Dow Jones time series ending in October 1987 and 1997, we
think it useful to compare the S\&P 500 time series ending in
October 1987 and 1997, to test for the robustness of the
log-periodic signal with respect to modifications and/or
alterations brought by different realization of the noise and of
the signal. Figures \ref{Fig:SP97DqHLombP1} to \ref{Fig:SP97PNf}
gives the results of the analysis of the $(H,q)$-derivative of
this time series. The results are highly consistent with those
obtained in the previous three time series.

\subsection{Time series ending with the Nasdaq Index, October 1987 crash}

\begin{figure}
\begin{center}
\epsfig{file=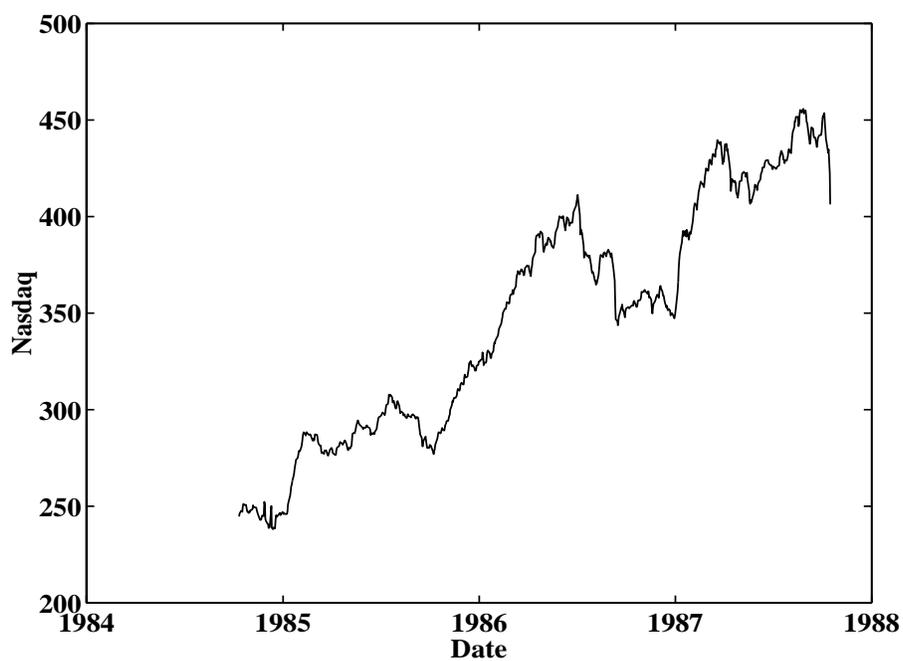,width=12cm, height=9cm}
\end{center}
\caption{Daily evolution of the Nasdaq Index from 11-Oct-1984
to the ``Black Monday'' on 19-Oct-1987.} \label{Fig:Nas87}
\end{figure}

\begin{figure}
\begin{center}
\epsfig{file=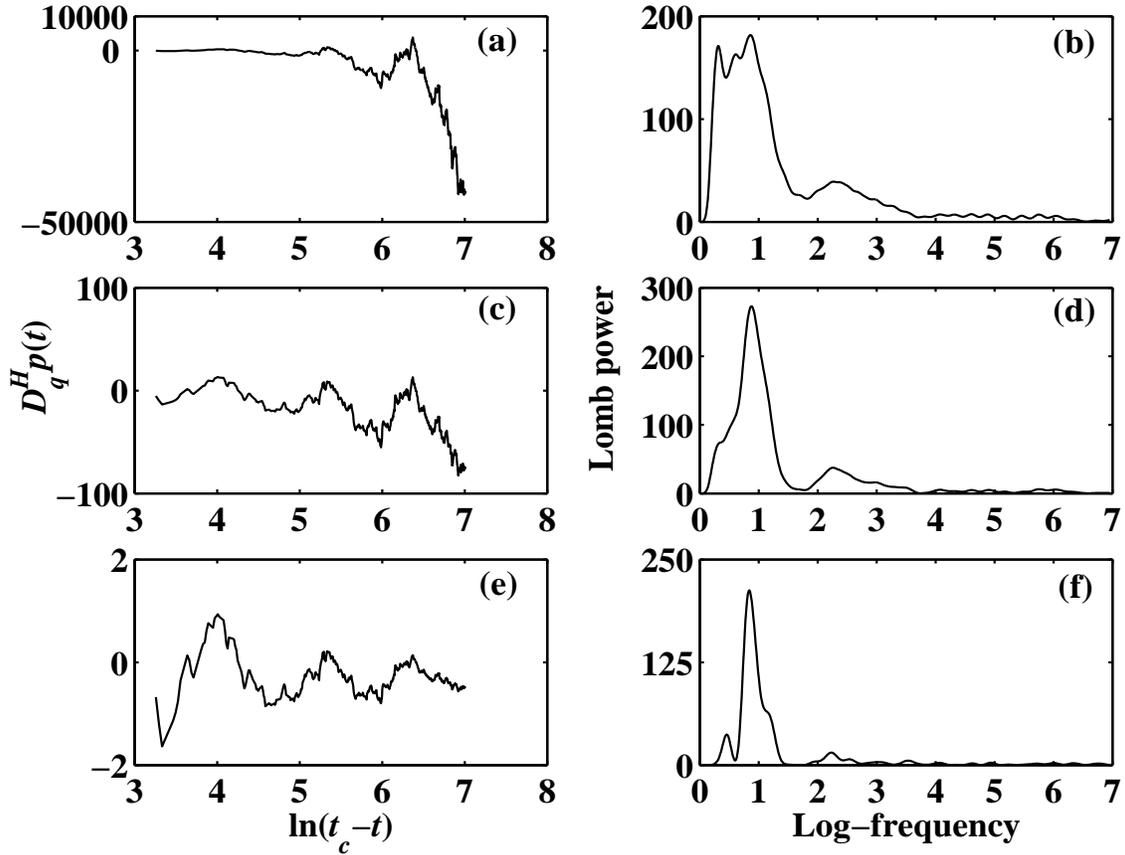, width=15cm, height=12cm}
\end{center}
\caption{The time evolution of the generalized $q$-derivative
(left panels) for the Nasdaq index time series ending with the
October 1987 Crash and their Lomb spectra (right panels) for fixed
$q = 0.5$: (a-b) $H = -0.9$; (c-d) $H = 0.1$; and (e-f) $H = 0.9$.
Three log-periodic oscillations of $D_q^Hp(t)$ in plots (c) and
(e) are clearly visible, while (a) exhibits a strong trend. The
amplitude of the log-periodic oscillations decreases with
increasing $H$ for fixed $q$. The envelops of the oscillations can
be fitted very well with Eq.~(\ref{Eq:Dm}). In the right panels,
we observe peaks at the following log-frequencies: $f_1 = 0.85$ in
(b), $f_1 = 0.88$ in (d) and $f_1 = 0.84$ in (f).}
\label{Fig:Nas87DqHLombP1}
\end{figure}

\begin{figure}
\begin{center}
\epsfig{file=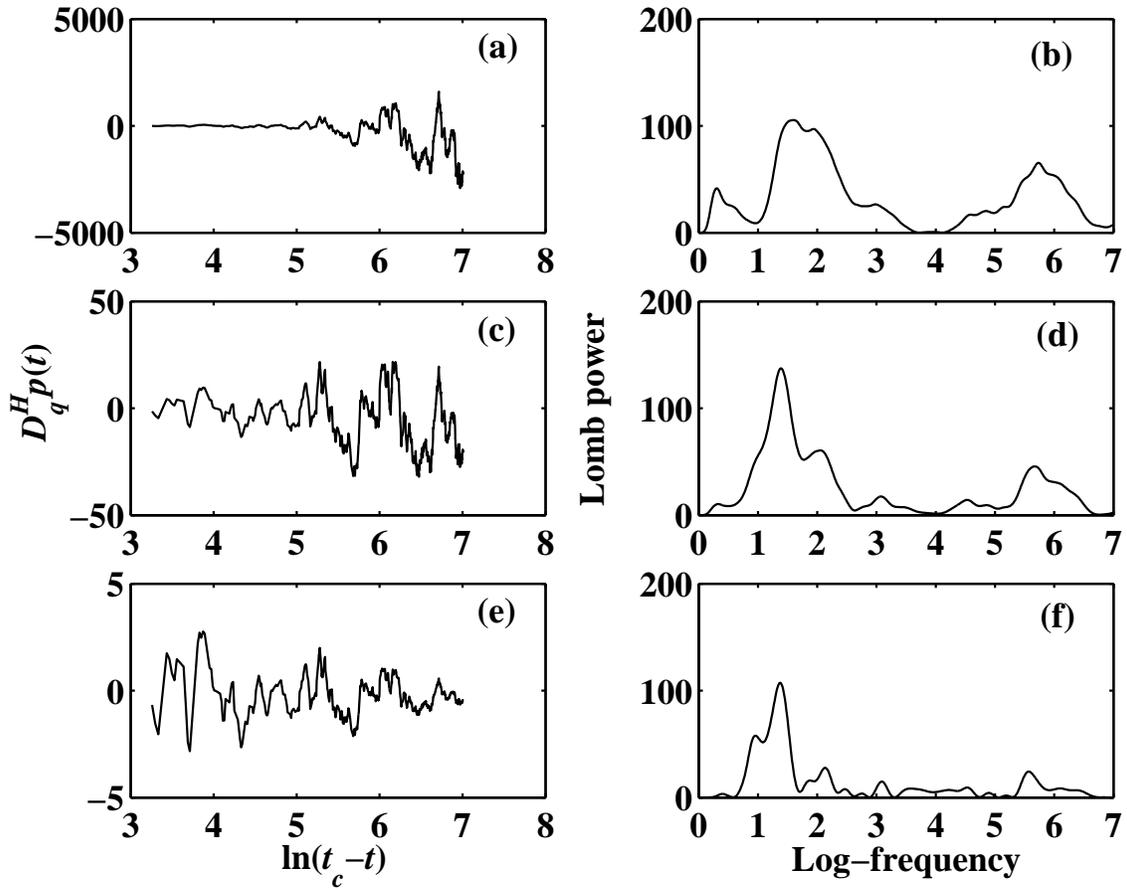, width=15cm, height=12cm}
\end{center}
\caption{Same as Fig.~\ref{Fig:Nas87DqHLombP1} for $q = 0.9$:
(a-b) $H = -0.9$; (c-d) $H = 0.1$; and (e-f) $H = 0.9$. The
log-periodic structures of $D_q^Hp(t)$ in plots (a), (c) and (e)
are clearly visible. The two highest Lomb peaks lie at $f_2 =
1.60$ in (b), $f_2 = 1.39$ in (d) and $f_2 = 1.37$ in (f).}
\label{Fig:Nas87DqHLombP2}
\end{figure}

\begin{figure}
\begin{center}
\epsfig{file=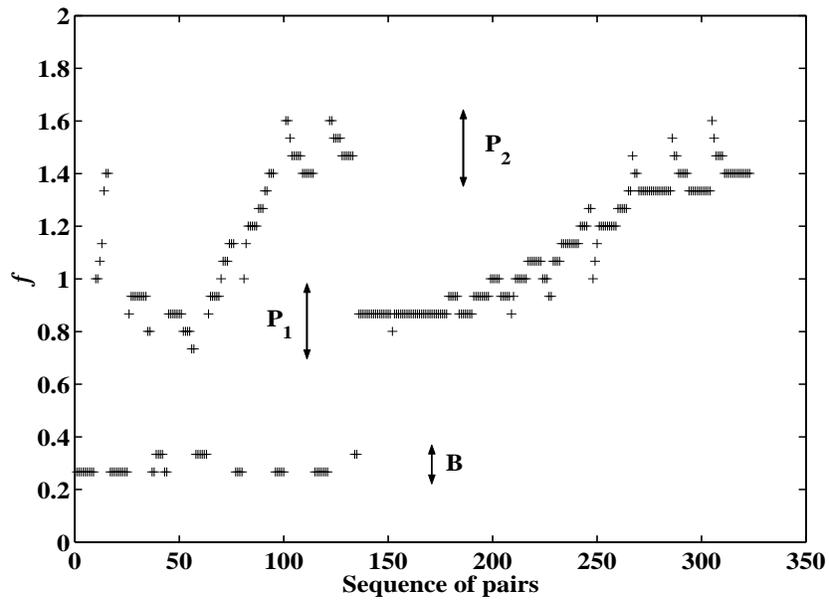, width=11cm, height=8cm}
\end{center}
\caption{Same as figure \ref{Fig:Dow87Pat} for the
Nasdaq time series ending at
the October 1987 crash. Three clusters can be observed:
${\mathbf{P_1}}$ with $f_1 = 0.88 \pm 0.04$,
${\mathbf{P_2}}$ with $f_2 = 1.45 \pm 0.06$ and
${\mathbf{B}}$ with $f_B \approx 0.21$ (which corresponds
as in other previously analyzed time series to approximately
one oscillation). Compared to previous time series, the
clusters are less clearly defined with a continuous range of
log-frequencies joining clusters ${\mathbf{P_1}}$ and
${\mathbf{P_2}}$.
}
\label{Fig:Nas87Pat}
\end{figure}

\begin{figure}
\begin{center}
\epsfig{file=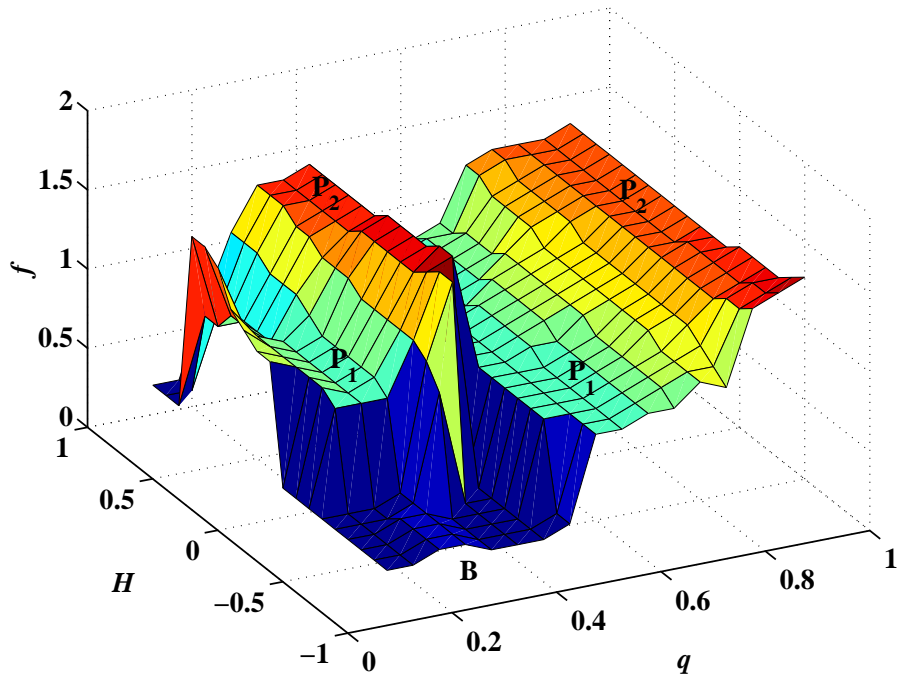, width=12cm, height=9cm}
\end{center}
\caption{Dependence of the log-frequency $f(H,q)$ of the
most significant peak in each Lomb periodogram of the
$(H,q)$-derivative for the Nasdaq index time series ending
at the October 1987 crash. One can observe two types of
platforms ${\mathbf{P_1}}$ and ${\mathbf{P_2}}$ with $f_1 = 0.88
\pm 0.04$ and $f_2 = 1.45 \pm 0.06$.} \label{Fig:Nas87f}
\end{figure}

\begin{figure}
\begin{center}
\epsfig{file=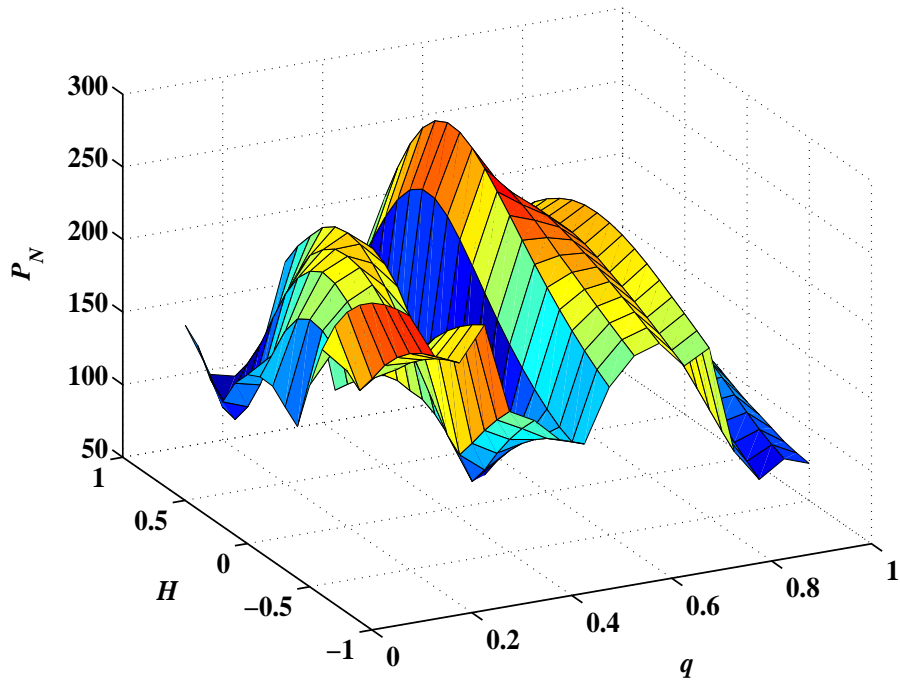, width=12cm, height=9cm}
\end{center}
\caption{Dependence of the highest peak $P_N(H,q)$ in each Lomb
periodogram of the $(H,q)$-derivative for the Nasdaq index time
series ending at the October 1987 crash.} \label{Fig:Nas87PN}
\end{figure}

\begin{figure}
\begin{center}
\epsfig{file=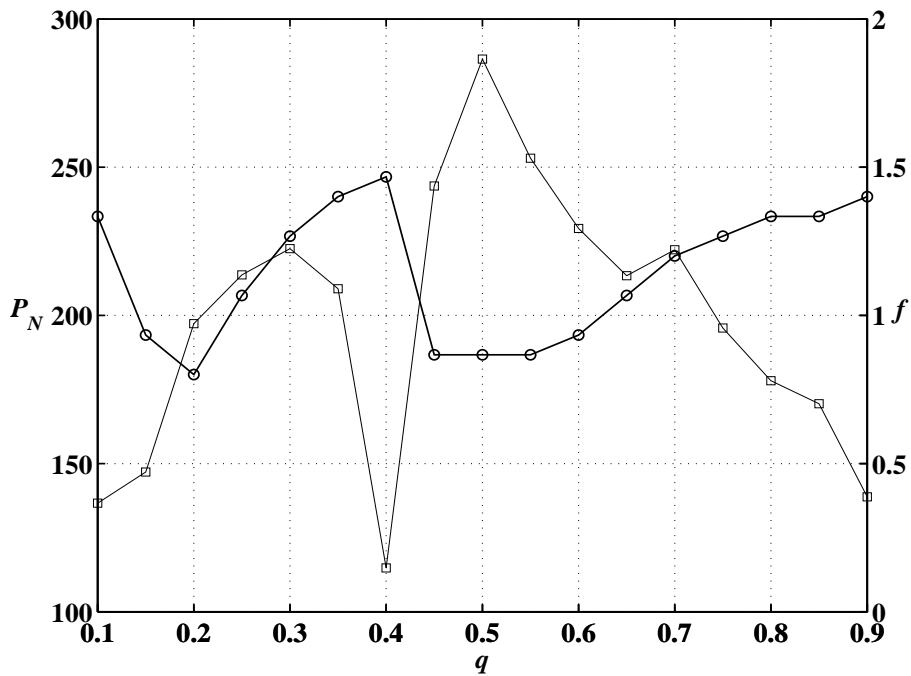, width=12cm, height=9cm}
\end{center}
\caption{Dependence wit $q$ for fixed $H=0.4$ of the highest peak
$P_N(\hat{H},q)$ of the Lomb spectrum (thin line marked with squares)
with scales given by the left vertical axis and of the
associated log-frequencies
$f(\hat{H},q)$ shown as the thick line marked with circles with scales
given by the right vertical axis, for the Nasdaq index
time series ending with the October 1987 crash. }
\label{Fig:Nas87PNf}
\end{figure}

Figure \ref{Fig:Nas87}  shows the daily evolution of the Nasdaq
Index from 11-Oct-1984 to the ``Black Monday'' on 19-Oct-1987,
which started the crash. Figures \ref{Fig:Nas87DqHLombP1} to
\ref{Fig:Nas87PNf} show the results of our analysis, which
essentially recovers similar results as for the previous time
series. The main difference is the weaker evidence of clustering
into three clusters ${\mathbf{P_1}}$ with $f_1 = 0.88 \pm 0.04$,
${\mathbf{P_2}}$ with $f_2 = 1.45 \pm 0.06$ and ${\mathbf{B}}$
with $f_B \approx 0.21$ (which corresponds as in other previously
analyzed time series to approximately one oscillation).

\subsection{Time series ending with the Nasdaq Index, October 1997
strong correction}

\begin{figure}
\begin{center}
\epsfig{file=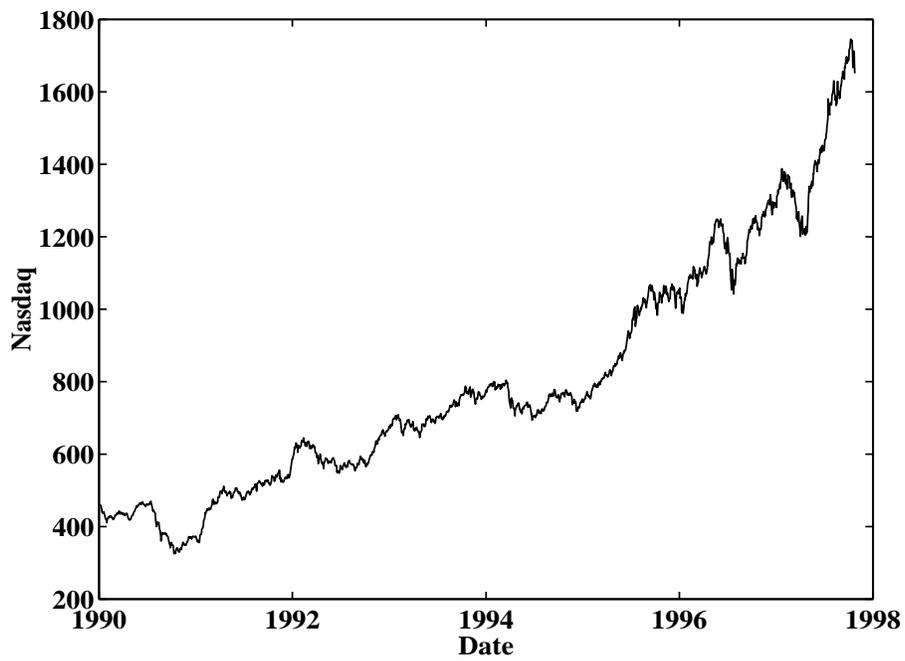,width=12cm, height=9cm}
\end{center}
\caption{Daily evolution of the Nasdaq Index from 02-Jan-1990
to 27-Oct-1997.} \label{Fig:Nas97}
\end{figure}

\begin{figure}
\begin{center}
\epsfig{file=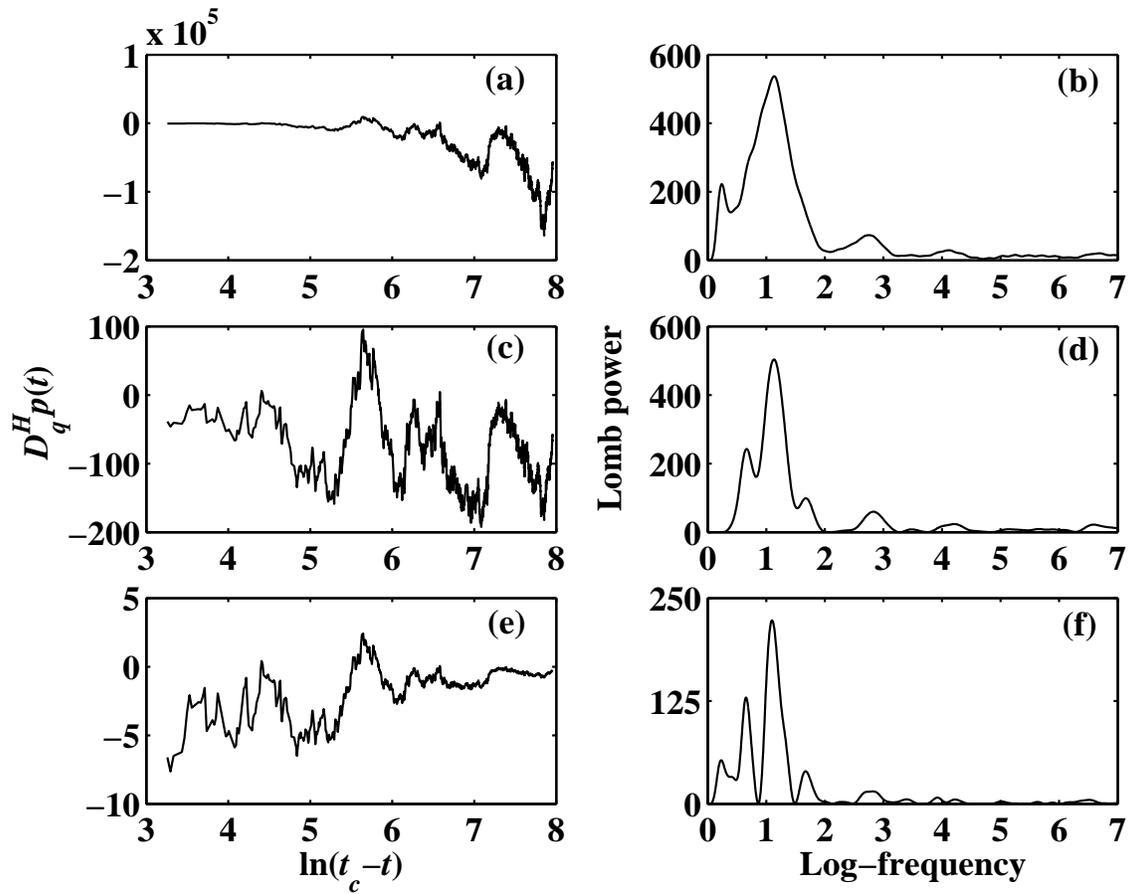, width=15cm, height=12cm}
\end{center}
\caption{Time evolution of the generalized $q$-derivative (left
panels) for the Nasdaq index ending with the October 1997
correction and their Lomb periodograms (right panels) for fixed $q
= 0.65$: (a-b) $H = -0.9$; (c-d) $H = 0.1$; and (e-f) $H = 0.9$.
The Lomb peaks at $f_1 = 1.14$ in (b), $f_1 = 1.13$ in (d) and
$f_1 = 1.10$ in (f) are very significant.}
\label{Fig:Nas97DqHLombP1}
\end{figure}

\begin{figure}
\begin{center}
\epsfig{file=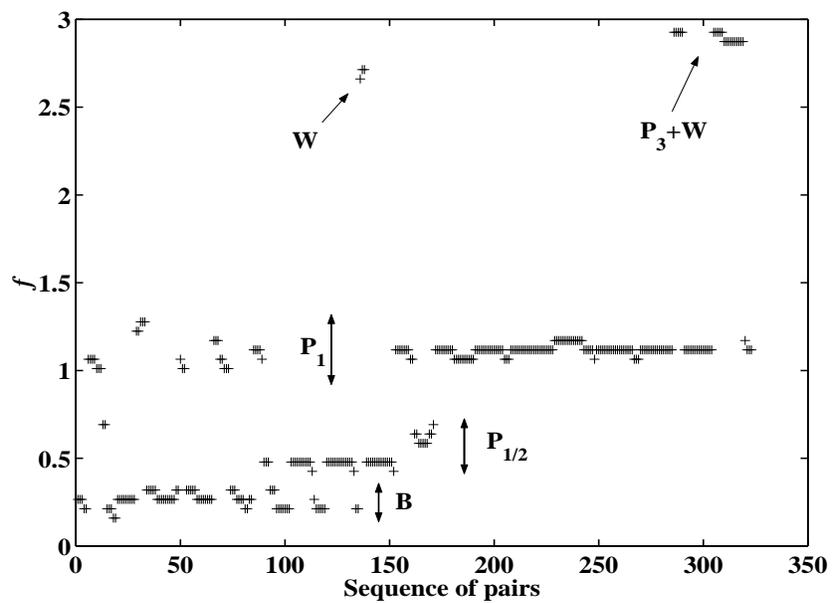, width=11cm, height=8cm}
\end{center}
\caption{Same as figure \ref{Fig:Dow87Pat} for the
Nasdaq time series ending at
the October 1997 correction. Several clusters can be observed:
  ${\mathbf{P_1}}$ with log-frequency $f_1 = 1.11 \pm
0.04$, ${\mathbf{P_{1/2}}}$ with $f_{1/2} = 0.51 \pm 0.07$,
${\mathbf{W}}+{\mathbf{P_3}}$ with $f_3 = 2.90 \pm 0.03$ and
${\mathbf{B}}$ with $f_B = 0.21$. See figure \ref{Fig:Nas97f}
for a better definition of the clusters.}
\label{Fig:Nas97Pat}
\end{figure}

\begin{figure}
\begin{center}
\epsfig{file=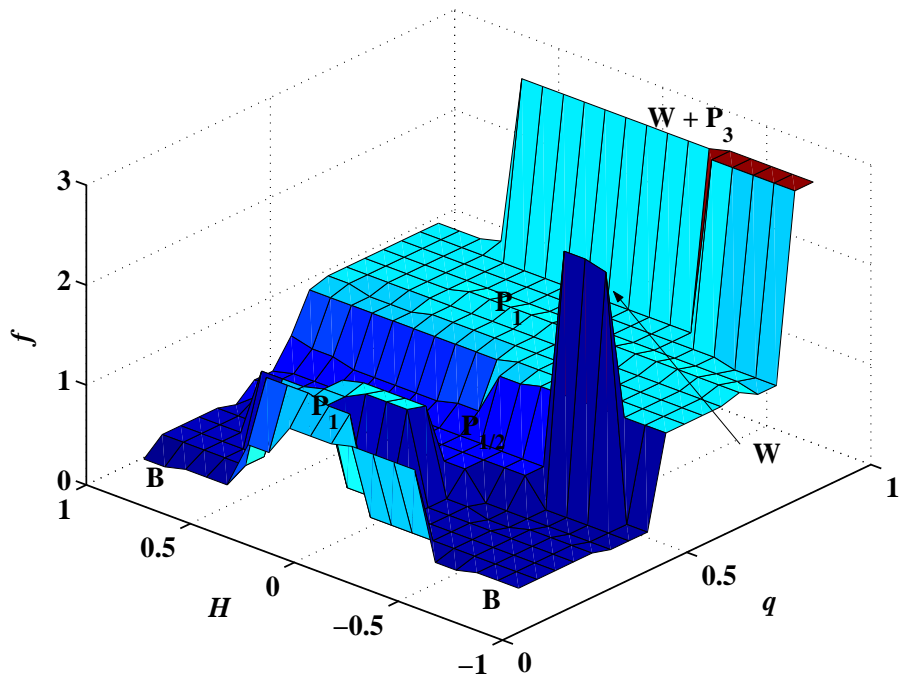, width=12cm, height=9cm}
\end{center}
\caption{Dependence of the log-frequency $f(H,q)$ of the
most significant peak in each Lomb periodogram of the
$(H,q)$-derivative for the Nasdaq index time series ending
with the October 1997 crash. One can identify several structures:
platform ${\mathbf{P_1}}$ with log-frequency
$f_1 = 1.11 \pm 0.04$, ${\mathbf{P_{1/2}}}$ with $f_{1/2} = 0.51 \pm
0.07$, ${\mathbf{W}}+{\mathbf{P_3}}$ with $f_3 = 2.90 \pm 0.03$.}
\label{Fig:Nas97f}
\end{figure}

\begin{figure}
\begin{center}
\epsfig{file=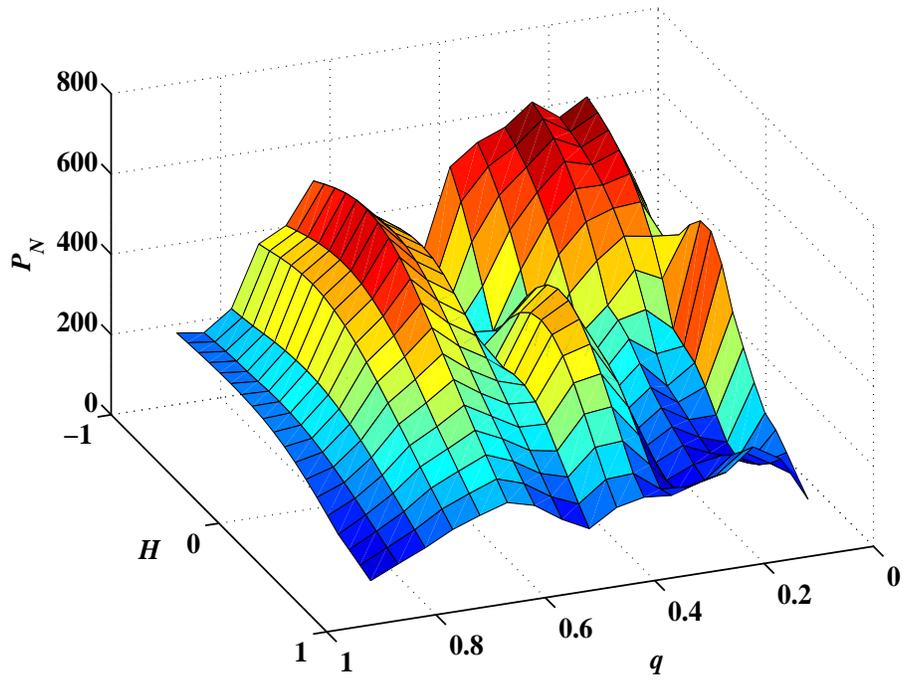, width=12cm, height=9cm}
\end{center}
\caption{Dependence of the highest peak $P_N(H,q)$ in each Lomb
periodogram of the $(H,q)$-derivative for the Nasdaq 1997 crash.}
\label{Fig:Nas97PN}
\end{figure}

\begin{figure}
\begin{center}
\epsfig{file=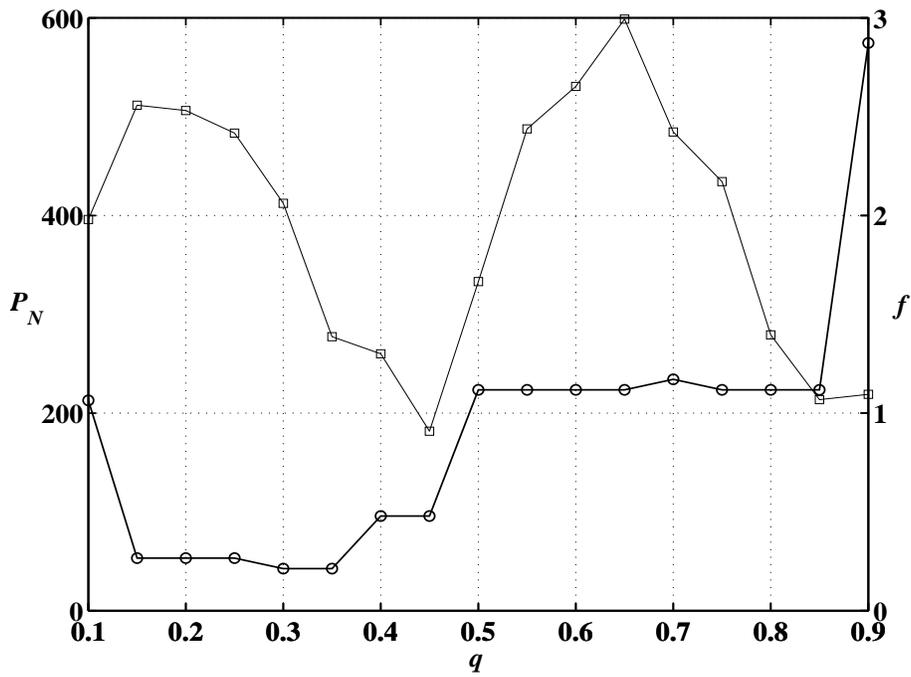, width=12cm, height=9cm}
\end{center}
\caption{Dependence with $q$ for fixed $H=-0.4$ of the highest peak
$P_N(\hat{H},q)$ shown as a thin line marked with squares with
left vertical axis and of the associated log-frequencies
$f(\hat{H},q)$ shown as the thick line marked with circles with right
vertical axis, for the Nasdaq index time series ending
at the October 1997 correction.} \label{Fig:Nas97PNf}
\end{figure}

Figure \ref{Fig:Nas97} shows the daily evolution of the Nasdaq
Index from 02-Jan-1990 to 27-Oct-1997, at which a strong
correction occurred. Figures \ref{Fig:Nas97DqHLombP1} to
\ref{Fig:Nas97PNf} show the results of our analysis of this time
series, which are very similar to previous ones.

\subsection{Time series ending with the Nasdaq Index, April 2000 crash}

\begin{figure}
\begin{center}
\epsfig{file=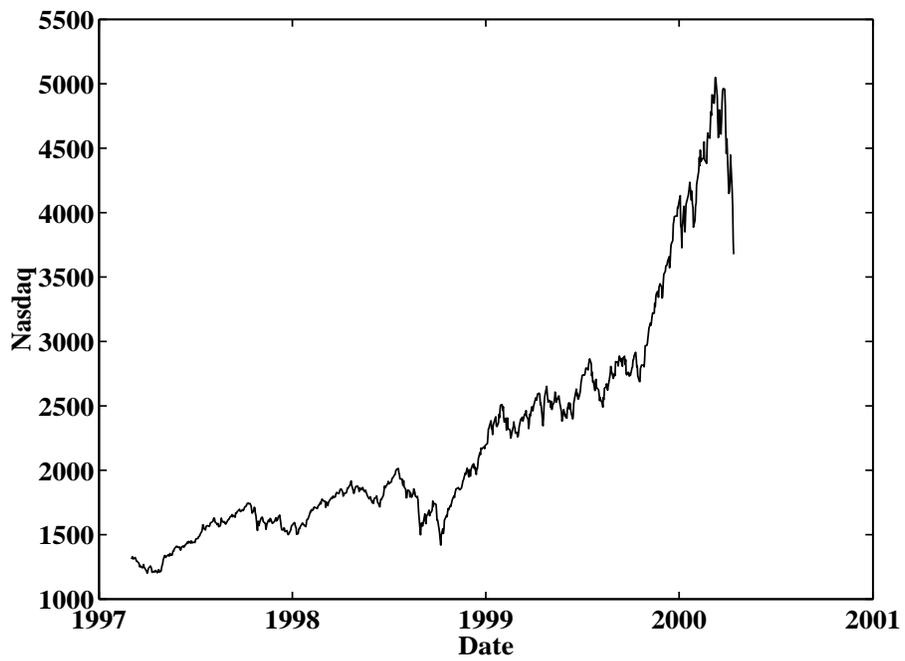,width=12cm, height=9cm}
\end{center}
\caption{Daily evolution of the Nasdaq Index from 03-Mar-1997
to 14-Apr-2000.} \label{Fig:Nas00}
\end{figure}

\begin{figure}
\begin{center}
\epsfig{file=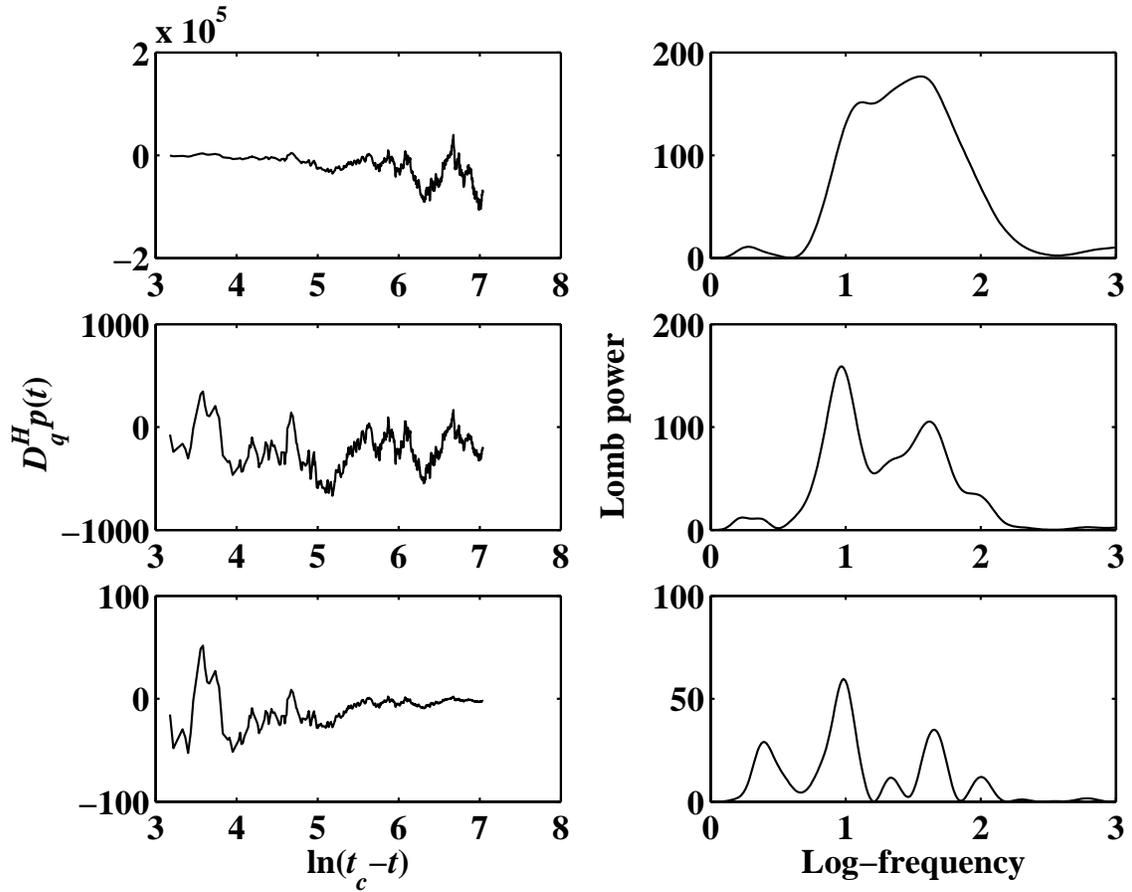, width=15cm, height=12cm}
\end{center}
\caption{Time evolution of the generalized $q$-derivative (left
panels) for the Nasdaq index time series ending with the
April 2000 crash and their Lomb periodograms
(right panels) for fixed $q = 0.7$: (a-b) $H = -0.9$; (c-d) $H =
0.1$; and (e-f) $H = 0.9$.} \label{Fig:Nas00DqHLombP1}
\end{figure}

\begin{figure}
\begin{center}
\epsfig{file=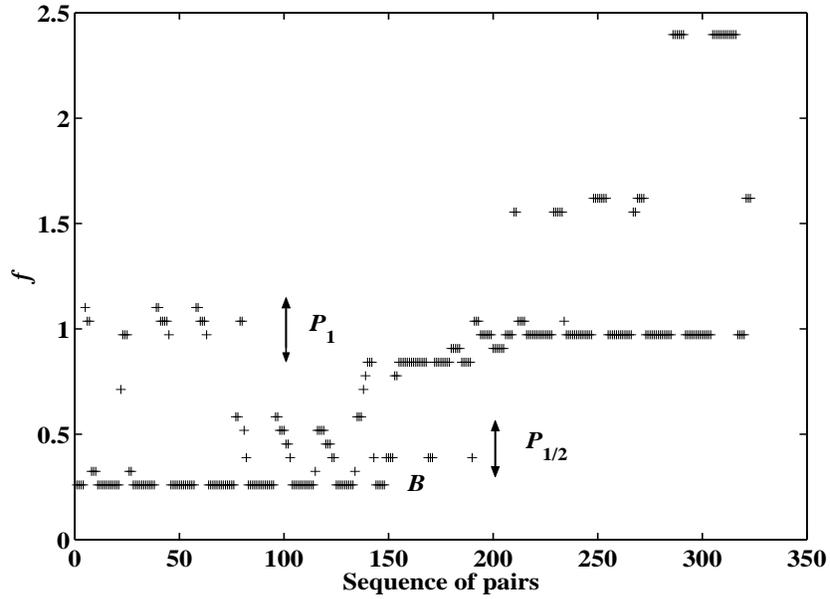, width=11cm, height=8cm}
\end{center}
\caption{Same as figure \ref{Fig:Dow87Pat} for the
Nasdaq time series ending at
the April 2000 crash. Several clusters can be observed:
${\mathbf{P_1}}$ with log-frequencies $f_1 = 0.98 \pm
0.04$, ${\mathbf{P_{1/2}}}$ with $f_{1/2} = 0.48 \pm 0.13$
and ${\mathbf{B}}$ with $f_B = 0.25$. Two small clusters are
also observed whose significance is not obvious.}
\label{Fig:Nas00Pat}
\end{figure}

\begin{figure}
\begin{center}
\epsfig{file=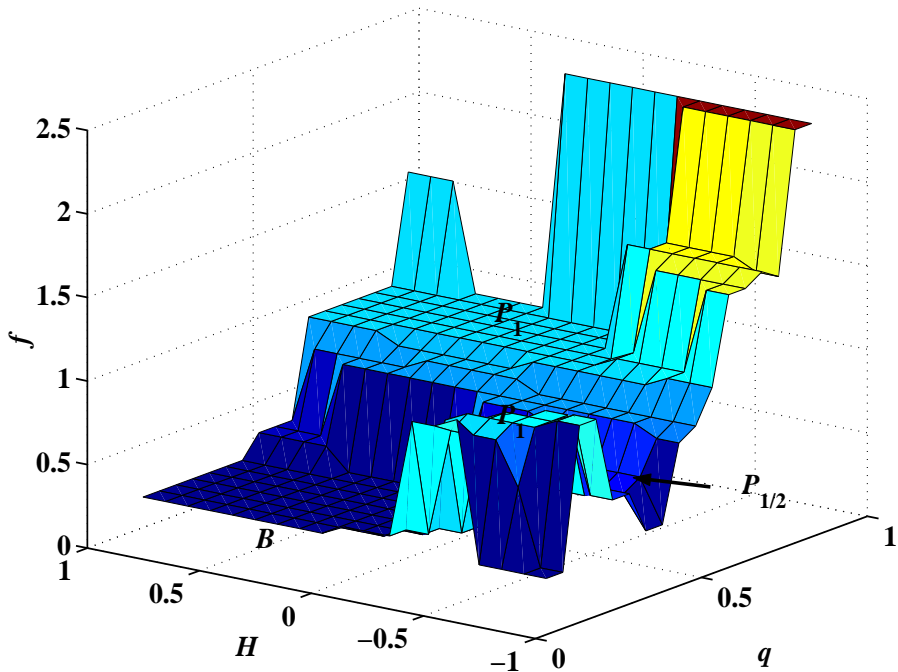, width=12cm, height=9cm}
\end{center}
\caption{Dependence of the log-frequency $f(H,q)$ of the
most significant peak in each Lomb periodogram of the
$(H,q)$-derivative for the Nasdaq index time series ending
with the April 2000 crash. The two main plateaux are
${\mathbf{P_1}}$ with log-frequency $f_1 = 0.98 \pm 0.04$,
${\mathbf{P_{1/2}}}$ with $f_{1/2} = 0.48 \pm 0.13$.}
  \label{Fig:Nas00f}
\end{figure}

\begin{figure}
\begin{center}
\epsfig{file=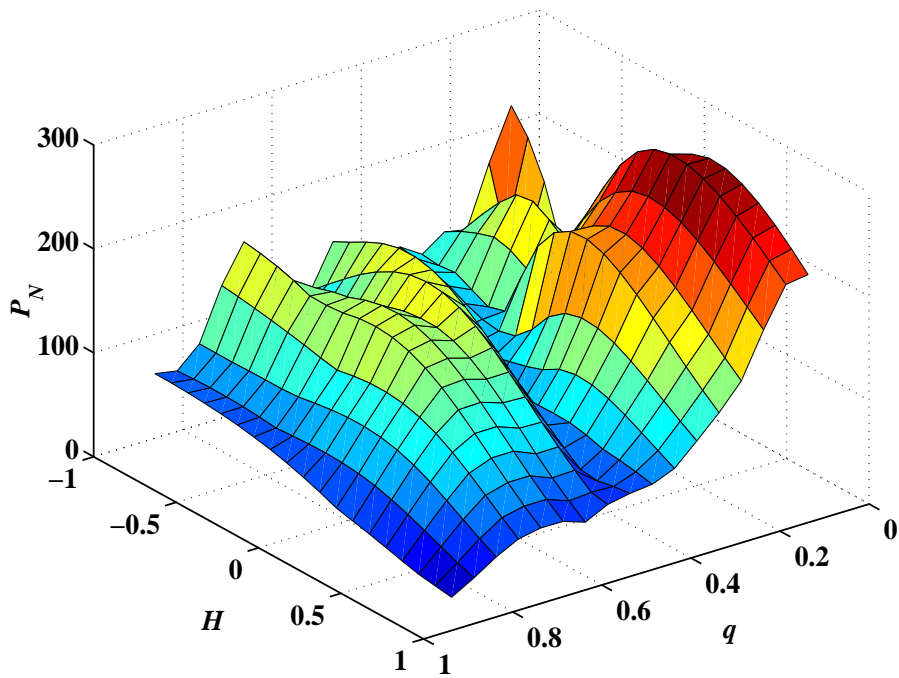, width=12cm, height=9cm}
\end{center}
\caption{Dependence of the highest peak $P_N(H,q)$ in each Lomb
periodogram of the $(H,q)$-derivative for the Nasdaq index time
series ending with the April 2000 crash.}
\label{Fig:Nas00PN}
\end{figure}

\begin{figure}
\begin{center}
\epsfig{file=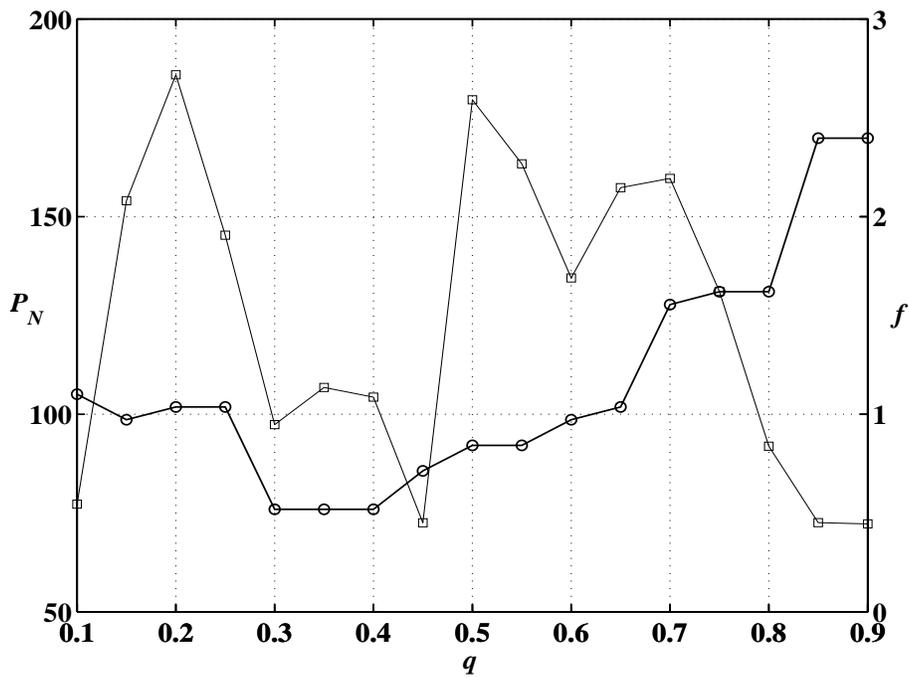, width=12cm, height=9cm}
\end{center}
\caption{Dependence with $q$ for fixed $H=-0.5$ of the highest peak
$P_N(\hat{H},q)$ shown as the thin line marked with squares
with the left vertical axis and of the associated log-frequencies
$f(\hat{H},q)$ shown as the thick line
marked with circles with the right vertical axis, for the
Nasdaq index time series ending with the April 2000 crash.}
\label{Fig:Nas00PNf}
\end{figure}

Figure \ref{Fig:Nas00} shows the time evolution of the daily
quotation of the Nasdaq Index from 03-Mar-1997 to 14-Apr-2000.
This crash corresponds to the burst of the so-called ``new
economy'' bubbles that has spurred in the few preceding years. We
refer to \cite{Nasdaq,bookcrash} for a detailed discussion of this
crash and a thorough analysis of its log-periodic signature using
a parametric approach as well as the non-parametric Lomb spectral
analysis of the residuals to the fit of the Nasdaq index time
series to a power law. Figures \ref{Fig:Nas00DqHLombP1} to
\ref{Fig:Nas00PNf} show the results of our analysis, which are
very similar to previous time series.

\section{Summary and discussion of the $(H,q)$-analysis}

By comparing the log-frequency $f(H,q)$ associated with the
highest peak $P_N(H,q)$ of the Lomb spectrum of the
$(H,q)$-derivative for the seven financial indices reported above,
we find striking similarities, in particular in the shape and
values of the plateaux and clusters of log-frequencies. Some of
these similarities can be attributed to the fact that all these
time series are not independent, especially those covering the
same time spans which use indices with overlapping stocks
\footnote{The DJIA is an index of 30 ``blue-chip'' US stocks. It
is the oldest continuing US market index. It is called an
``average'' because it was originally computed by adding up stock
prices and dividing by the number of stocks (the very first
average price of industrial stocks, on May 26, 1896, was 40.94)
and should ideally represent a correct measure of the state of the
economy. The methodology remains the same today, but the divisor
has been changed to preserve historical continuity. The standard
\& Poor 500 index is a capitalization weighted average of 500 of
the major stocks on the US market. Thus the S\&P 500 includes the
30 stocks constituting the Dow Jones index and it is natural to
expect strong correlations between the two, especially at times of
strong herding as in speculative bubbles.}. However, the found
universality of the fundamental log-frequency is quite remarkable
as it seems independent of the specific realization of the noise
decorating the log-periodic patterns, as proxies by the different
indices for the same time periods, and is robust over the three
time periods that have been investigated.

We summarize in Table \ref{Tb:Sum} all the results presented in
previous figures by listing the most significant log-frequencies
for each of the seven time series. Apart from the Nasdaq time
series ending on Oct. 1987, all the other 6 time series give a
``fundamental'' log-frequency which is remarkably universal and
compatible with a universal value close to $1$. This value
corresponds to a preferred scaling ratio between successive maxima
of the oscillations equal to $\lambda = e^{1/f} \approx 2.7$. One
can also observe the presence of harmonics and of the first
sub-harmonic of this fundamental log-frequency, which reinforce
the evidence for the existence of a discrete scaling invariance:
indeed, the sub-harmonic corresponds to $\lambda^2$, that is, to
two steps of the discrete hierarchy; as for the harmonics, any
nonlinearity in the price dynamics will produce them from a
fundamental log-frequency. Taking into account of the
log-frequencies $f_1$, $f_2$ and $f_3$, excluding the Nasdaq 1987
time series, and assuming that all these log-frequencies derive
from a universal fundamental value $f$, our best estimate for $f$
is then $f = 1.02 \pm 0.05$. The corresponding preferred scaling
ratio is then $\lambda = 2.67 \pm 0.12$, which is in excellent
agreement with that from other analyses \cite{Vande,SorJoh01}. The
standard deviation of $\lambda$ is evaluated by the formulae
$\sigma_\lambda = {e^{1/f} } \sigma_f / f^2$. Our analysis also
suggest that the most sensitive range for $q$ lies in the interval
$q \in [0.6, 0.85]$.

The Nasdaq 87 time series seems to belong to a different class
with a significantly smaller fundamental frequency $f_1$. This may
be due to the fact that the Nasdaq was more like an ``emerging''
market before the October 1987 Crash, and thus shared the property
of having larger fluctuations, as found in a recent systematic
study of other emerging markets \cite{emergent}.

\begin{table}
\begin{center}
\begin{tabular}{|c|c|c|c|c|c|c|c|}
\hline
Crashes&$f_1$&$f_2$&$f_3$&$f_{1/2}$&$(\hat{H},\hat{q})$&$\hat{f}$&$\hat{P}_N$\\\hline
Dow Jones 87 &$1.04\pm 0.11$&$1.98 \pm 0.07$&/&/&$(-0.7,
0.60)$&0.96&580\\\hline Dow Jones 97 & $1.04 \pm 0.05$ & $2.34$ &
/& $0.59\pm 0.04$ & $(-0.8, 0.60)$ &1.06 &503\\\hline S\&P 500 87
& $1.00\pm0.11$ & $1.97 \pm 0.06$ & /&/&$(-0.3,0.80)$&1.97&516
\\\hline S\&P 500 97 & $1.06\pm0.04$ &/&/& $0.61\pm 0.05$
&$(-0.5,0.65)$&1.06 &652 \\\hline Nasdaq 87&$0.88\pm 0.04$& $1.45
\pm 0.06$ &/&/&$(\,\,\,\,0.4, 0.50)$&0.87&286 \\\hline Nasdaq 97 &
$1.11 \pm 0.04$ & /& $2.90 \pm 0.03$&$0.51 \pm 0.07$&$(-0.4,
0.65)$&1.12&599\\\hline Nasdaq 00 & $0.98 \pm 0.04$ & /& / & $0.48
\pm 0.13$&$(-0.5, 0.20)$&1.04&186\\\hline
\end{tabular}
\end{center}
\caption{Synthesis of the results of the analysis of seven price
time series using the $(H,q)$-derivative. The periods of the
analyzed data sets are: from 02-Jan-1980 to 19-Oct-1987 for the
Dow Jones and S\&P 500 indices, from 11-Oct-1984 to 19-Oct-1987
for the Nasdaq index, from 02-Jan-1990 to 27-Oct-1997 for the Dow
Jones, S\&P 500 and Nasdaq Indices, and from 03-Mar-1997 to
14-Apr-2000 for the Nasdaq index. The columns $f_1$, $f_2$, $f_3$
and $f_{1/2}$ are respectively the fundamental log-frequency, its
second and third harmonics and its first sub-harmonic. The columns
$(\hat{H}, \hat{q})$, $\hat{f}$ and $\hat{P}_N$ are the optimal
pairs and their corresponding log-frequency and their Lomb peak
height.} \label{Tb:Sum}
\end{table}

\section{Analysis of log-periodicity using the Hilbert transform}

\subsection{Definition and useful properties of the Hilbert transform}

The Hilbert transform (HT) of a signal $x(t)$ is defined by the
equation \be {\hat x}(s) = {1 \over \pi} \int_{-\infty}^{+\infty}
{x(t) \over s-t} dt~, \label{nglfld} \ee where the integral is the
Cauchy principal value. The reconstruction formula \be x(t) = -{1
\over \pi} \int_{-\infty}^{+\infty} {x(s) \over t-s} ds \ee
defines the inverse HT. Expression (\ref{nglfld}) shows that
${\hat x}(s) = (1/\pi t) \otimes x(t)$ is the convolution of
$(1/\pi t)$ with $x(t)$, which by taking the Fourier transform
implies that \be {\hat X}(\nu) = - j~{\rm sign}(\nu) X(\nu)~, \ee
where $j^2=-1$ and $X(\nu)$ (respectively ${\hat X}(\nu)$) is the
Fourier transform of $x(t)$ (respectively ${\hat x}(s)$). This
means that the HT produces a $-\pi/2$ radian phase shift for the
positive frequency components of the input $x(t)$, while the
amplitude does not change. A signal $x(t)$ and its HT ${\hat
x}(s)$ have the same amplitude spectrum, the same autocorrelation
function, they are orthogonal and the HT of ${\hat x}(s)$ is
$-x(t)$.

The HT is thus an ideal tool for detecting an arbitrary phase
function $P(t)$ in a signal of the form $x(t) =
A(t)\cos(\theta(t)) + n(t)$, where $n(t)$ is a noise with small
amplitude ($n \ll A$), and $A(t)$ is a slowly varying amplitude
modulation. The phase $\theta(t)$ is in principle recovered by the
formula \be \theta(t) = \tan^{-1}[{\hat x}(t)/x(t)]~. \ee However,
the phase obtained in this way consists in saw-tooth peaks that
drop sharply from $\pi$ down to $-\pi$. It is thus necessary to
unwrap the phase to obtain a smooth continuous phase.

\begin{figure}
\begin{center}
\epsfig{file=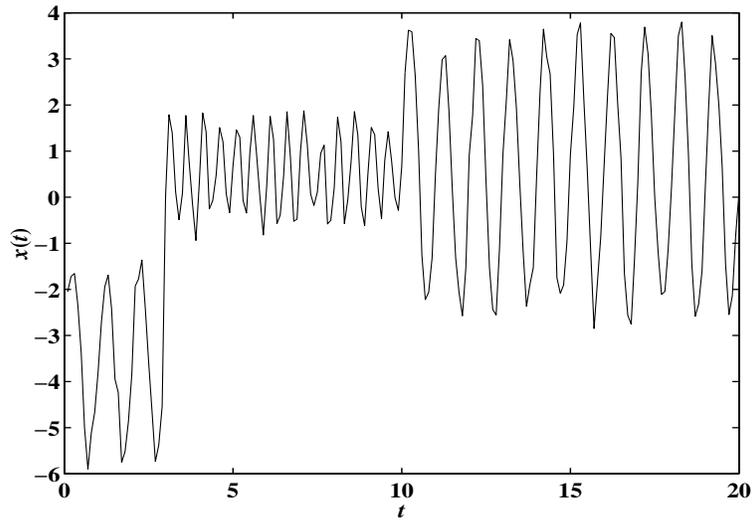,width=10cm, height=7cm}
\end{center}
\caption{An example of function (\ref{Eq:HT2xt}) exhibiting a
combination of an abrupt translation with phase velocity and
amplitude changes.} \label{Fig:HT2xt}
\end{figure}

\begin{figure}
\begin{center}
\epsfig{file=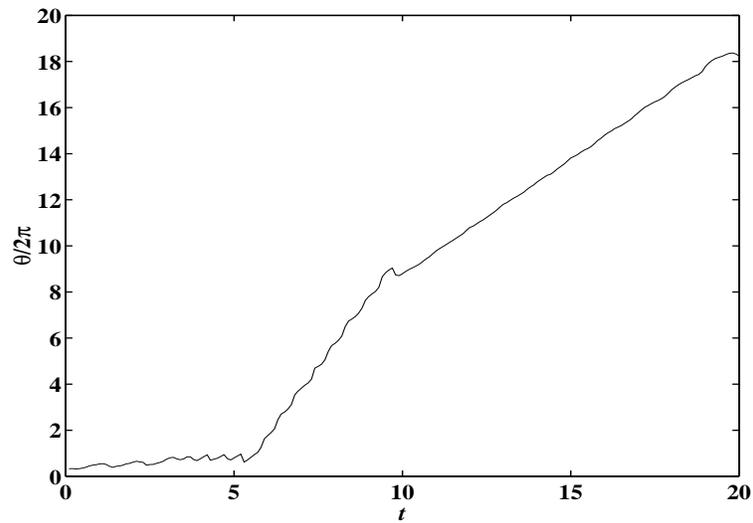,width=10cm, height=7cm}
\end{center}
\caption{Unwrapped phase of the function shown in
Fig.~\ref{Fig:HT2xt}.}
\label{Fig:HT2Th}
\end{figure}

An example is given in figure \ref{Fig:HT2xt} which shows the function
\begin{equation}
x(t) = \epsilon(t) + \left\{ {\begin{array}{*{20}c}
      2\sin(2\pi x)-4, & 0 < x \leq 3,  \\
      \sin(4\pi x), & 3 < x \leq 10, \\
      3\sin(2\pi x), & 10 <x \leq 20. \\
\end{array} } \right. \label{Eq:HT2xt}
\end{equation}
where $\epsilon(t)$ is normally distributed, which consists of a
high-frequency burst in a low-frequency background. The unwrapped
phase $\theta(t)$ is plotted in Fig.~\ref{Fig:HT2Th}. The abrupt
translation of $x(t)$ at $t=0.3$ leads to an incorrect
determination of the phase in this interval, and the corresponding
distortion also spreads in the high-frequency interval
approximately up to $t=5.5$. Otherwise, the abrupt change of
frequency is correctly detected at $t=10$ notwithstanding the
abrupt change in amplitude.

\subsection{Hilbert transform to financial time series}

\begin{figure}
\begin{center}
\epsfig{file=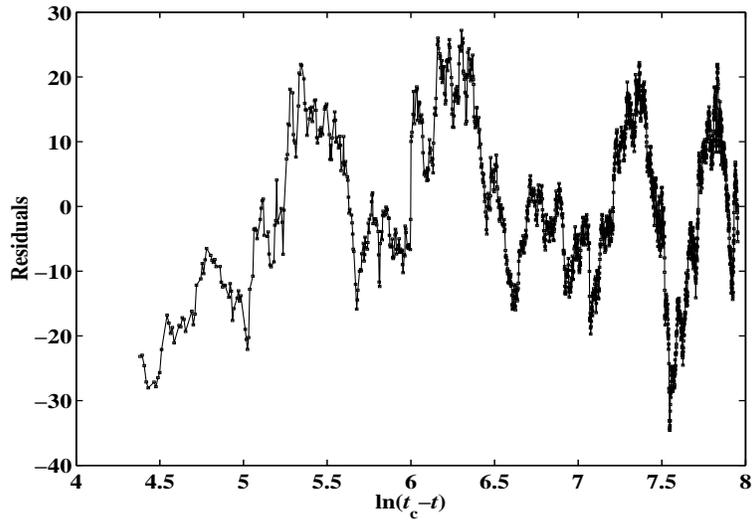,width=10cm, height=7cm}
\end{center}
\caption{Difference between the S\&P 500 price time series ending
with the October 1987 crash and the fit with $A-B\ln(t_c-t)$,
where $A$ and $B$ are two constants, as a function of
$\ln(t_c-t)$. $t_c$ is the time of the crash.}
  \label{Fig:SP5ResNN87}
\end{figure}

\begin{figure}
\begin{center}
\epsfig{file=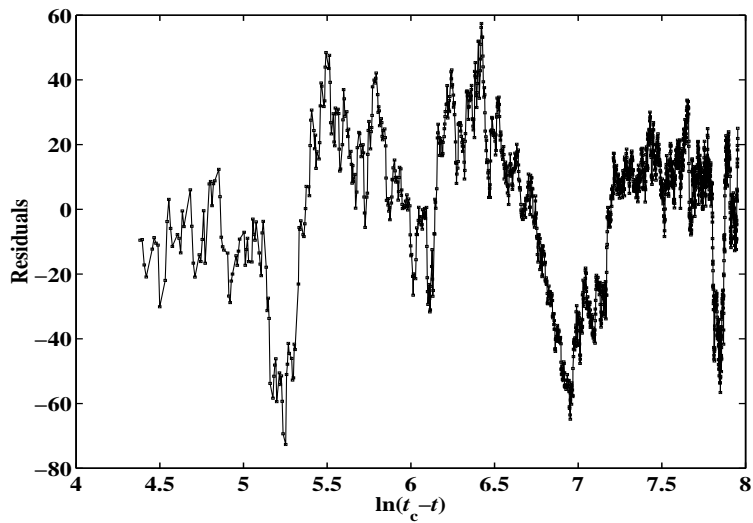,width=10cm, height=7cm}
\end{center}
\caption{Same as figure \ref{Fig:SP5ResNN87} for the S\&P 500
price time series ending with the October 1997
correction.}\label{Fig:SP5ResNN97}
\end{figure}

\begin{figure}
\begin{center}
\epsfig{file=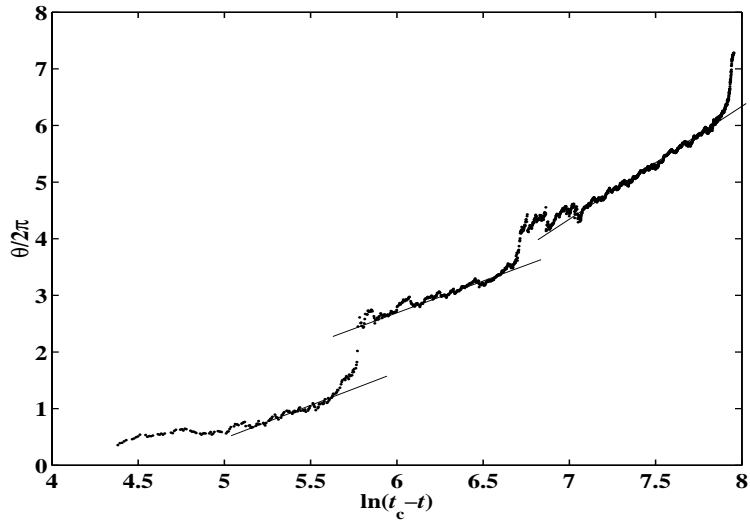,width=10cm, height=7cm}
\end{center}
\caption{Unwrapped phase $\theta$ extracted from the residuals in
Fig.~\ref{Fig:SP5ResNN87} using the Hilbert transform. The slopes
of the three linear segments are respectively the log-frequencies
$f = 0.9, f= 0.9$ and $f = 1.9$. This is in agreement with the
existence of a fundamental log-frequency and its harmonics. The
jumps in the phase are due to the fast oscillations of small
amplitudes in the interval $\ln(t_c-t) \in (5,7)$.}
\label{Fig:SP5ThNN87}
\end{figure}

\begin{figure}
\begin{center}
\epsfig{file=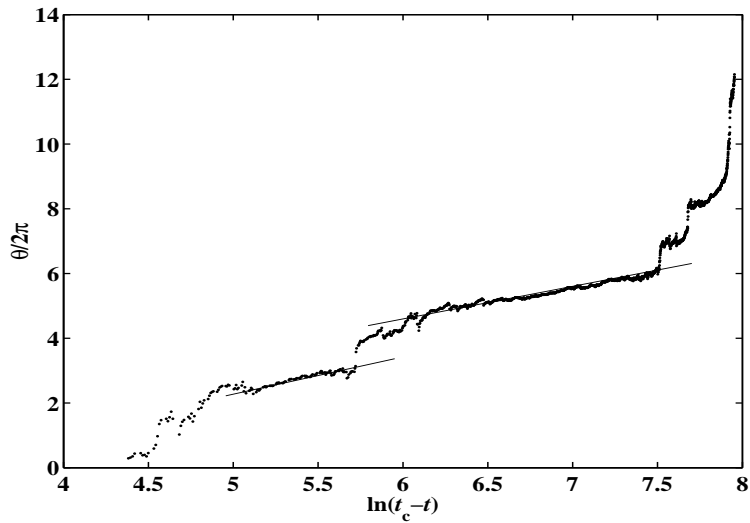,width=10cm, height=7cm}
\end{center}
\caption{Unwrapped phase $\theta$ extracted from the residuals
shown in Fig.~\ref{Fig:SP5ResNN97} using the Hilbert transform. Two parallel
linear segments are indicated by the straight lines whose slopes $\approx 1.0$
give the value of the log-frequency. The
jumps are caused by the fast fluctuations
in the second oscillation in Fig.~\ref{Fig:SP5ResNN97} near
$\ln(t_c-t) = 5.7$. The fluctuations in the fourth oscillation for
$\ln(t_c-t) > 7.4$ in Fig.~\ref{Fig:SP5ResNN97} completely spoil
the linear dependence of the unwrapped phase.}
\label{Fig:SP5ThNN97}
\end{figure}

Before applying the HT transform, we prepare the data (financial
time series) by transforming the time variable $t$ into $x \equiv
\ln (t_c -t)$, where $t_c$ is fixed as in the $(H,q)$-analysis to
the time of the realized crash. The new variable $\tau$ is such
that log-periodic oscillations in the variable $t$ become regular
periodic oscillations in the variable $x$. We then fit the price
time series as a linear function of $x$. The residuals of this fit
for the two S\&P 500 price time series ending with the October
1987 and the October 1997 correction are shown in figures
\ref{Fig:SP5ResNN87} and \ref{Fig:SP5ResNN97}. We then apply the
Hilbert transform to these residuals and obtain the corresponding
unwrapped phase in figures \ref{Fig:SP5ThNN87} and
\ref{Fig:SP5ThNN97}. Apart from jumps and distortions at the
boundaries due to fast oscillations and rapid changes of
amplitudes, the unwrapped phases provide a reliable determination
of the log-frequency in the vicinity of $f=1$, in agreement with
the results obtained above with the $(H,q)$-analysis.

\begin{figure}
\begin{center}
\epsfig{file=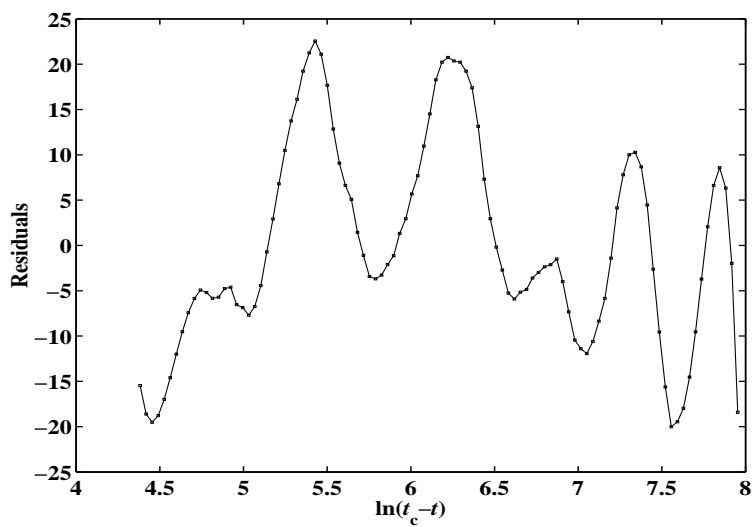,width=10cm, height=7cm}
\end{center}
\caption{The residuals for the October 1987 crash time of the S\&P
500 index from the smoothened data using spline interpolation and
then Savitzky-Golay filter. The 100 interpolation points are
evenly sampled in $\ln(t_c-t)$.} \label{Fig:SP5ResYY87}
\end{figure}

\begin{figure}
\begin{center}
\epsfig{file=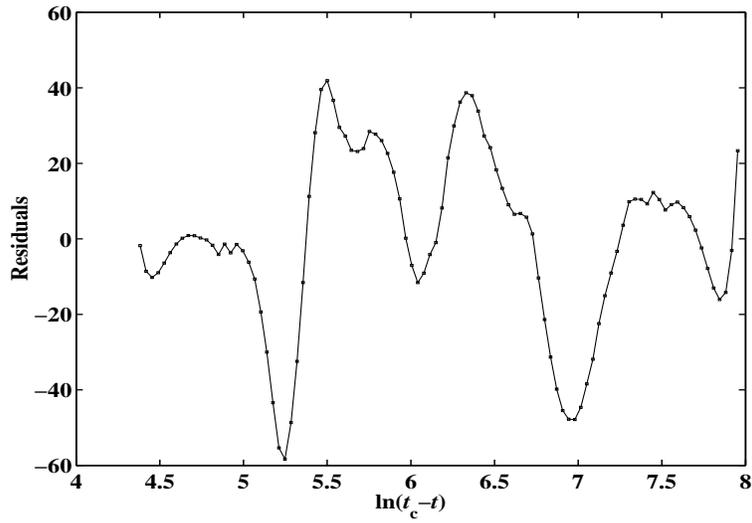,width=10cm, height=7cm}
\end{center}
\caption{Same as Fig. ~\ref{Fig:SP5ResYY87} but for the October
1997 correction of the S\&P 500 index.} \label{Fig:SP5ResYY97}
\end{figure}

\begin{figure}
\begin{center}
\epsfig{file=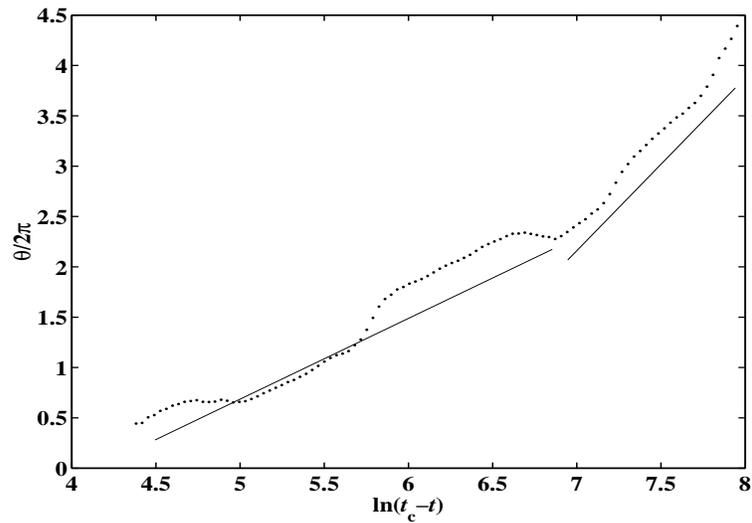,width=10cm, height=7cm}
\end{center}
\caption{The unwrapped phase $\theta$ extracted from the residuals
in Fig.~\ref{Fig:SP5ResYY87} using the Hilbert transform. Small
oscillations within one oscillation lead to the undulations in the
unwrapped phase. The two straight lines define log-frequencies
equal respectively to $f=0.85$ and $f=1.8$. }
\label{Fig:SP5ThYY87}
\end{figure}

\begin{figure}
\begin{center}
\epsfig{file=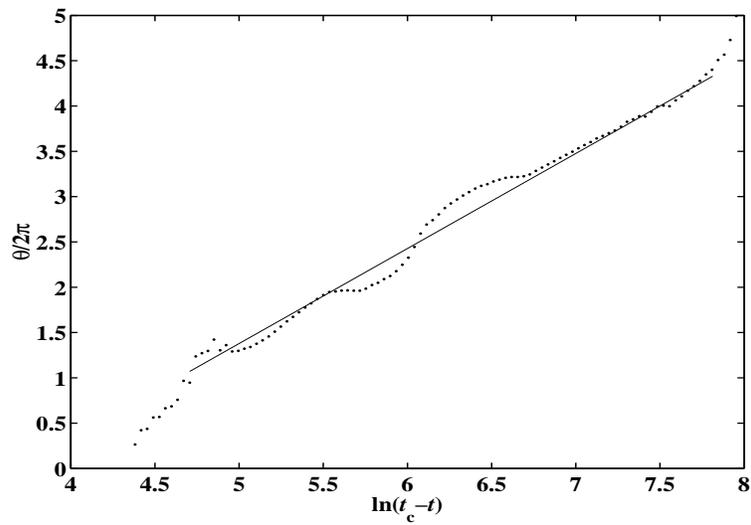,width=10cm, height=7cm}
\end{center}
\caption{The unwrapped phase $\theta$ extracted from the residuals
in Fig.~\ref{Fig:SP5ResYY97} using the Hilbert transform. The main
part of the phase can be fitted to a straight line of slope giving
the log-periodic frequency $f = 1.0$. Small oscillations within
one oscillation lead to the undulations in the unwrapped phase.}
\label{Fig:SP5ThYY97}
\end{figure}

As an attempt to improve the HT, we first smooth the financial
time series in two steps, by first performing a spline
interpolation with 100 evenly spaced sampling points in
$\ln(t_c-t)$ and then applying the Savitzky-Golay filter with
centered window of width 21 and six-order polynomials. Such
smoothing will alleviate some of the problems in the HT coming
from fast-varying amplitudes. The resulting residuals (that is,
filtered versions of the curves shown in figures
\ref{Fig:SP5ResNN87} and \ref{Fig:SP5ResNN97} are given in figures
\ref{Fig:SP5ResYY87} and \ref{Fig:SP5ResYY97}. The corresponding
unwrapped phases obtained by taking the HT of these data are shown
in figures \ref{Fig:SP5ThYY87} and \ref{Fig:SP5ThYY97}. The slope
of these unwrapped phases provide a good estimation of the
log-periodic frequency close to $1$, in very good agreement with
the $(H, q)$ analysis.

\section{Conclusion}
\label{s:conc}

In summary, we have applied two non-parametric analyses to
test the hypothesis that log-periodicity is present in financial
time series preceding crash or strong corrections. The $(H,q)$-derivative
and the Hilbert transform performed on the detrended price in the
$\ln(t_c-t)$ variable are both consistent with the existence of a strong
log-periodic signal with log-frequency $f \approx 1$.

\bigskip
{\bf Acknowledgments:} We are grateful to Vladilen Pisarenko for
constructive help in Hilbert transform. This work was partially
supported by the James S. Mc Donnell Foundation 21st century
scientist award/studying complex system.

\pagebreak

\clearpage


\clearpage


\clearpage


\clearpage


\clearpage


\clearpage


\clearpage


\clearpage


\clearpage

\end{document}